 \documentclass[final,3p,times,twocolumn,authoryear]{elsarticle}

\usepackage{amssymb}
\usepackage{natbib}
\usepackage{algorithm}
\usepackage{algpseudocode}
\usepackage{amsmath}
\usepackage{color}
\usepackage{xcolor}

\newcommand{\CASItD}{{\sc casi-3d}}
\newcommand{\CASItwoD}{{\sc casi-2d}}

\newcommand{\xd}[1]{{\textcolor{black}{#1}}}
\newenvironment{xdpar}{\par\color{black}}{\par}

\journal{VSI: ML methods in astronomy}

\begin{document}

\begin{frontmatter}



\title{Surveying Image Segmentation Approaches in Astronomy}


\author[label1]{Duo Xu}
\affiliation[label1]{organization={Department of Astronomy, University of Virginia},
            addressline={530 McCormick Rd}, 
            city={Charlottesville},
            postcode={22904-4235}, 
            state={VA},
            country={USA}}
            
\author[label2]{Ye Zhu}
\affiliation[label2]{organization={Department of Computer Science,Princeton University},
            addressline={35 Olden St}, 
            city={Princeton},
            postcode={08540}, 
            state={NJ},
            country={USA}}

\begin{abstract}
Image segmentation plays a critical role in unlocking the mysteries of the universe, providing astronomers with a clearer perspective on celestial objects within complex astronomical images and data cubes. Manual segmentation, while traditional, is not only time-consuming but also susceptible to biases introduced by human intervention. As a result, automated segmentation methods have become essential for achieving robust and consistent results in astronomical studies. This review begins by summarizing traditional and classical segmentation methods widely used in astronomical tasks. Despite the significant improvements these methods have brought to segmentation outcomes, they fail to meet astronomers' expectations, requiring additional human correction, further intensifying the labor-intensive nature of the segmentation process. The review then focuses on the transformative impact of machine learning, particularly deep learning, on segmentation tasks in astronomy. It introduces state-of-the-art machine learning approaches, highlighting their applications and the remarkable advancements they bring to segmentation accuracy in both astronomical images and data cubes. As the field of machine learning continues to evolve rapidly, it is anticipated that astronomers will increasingly leverage these sophisticated techniques to enhance segmentation tasks in their research projects. In essence, this review serves as a comprehensive guide to the evolution of segmentation methods in astronomy, emphasizing the transition from classical approaches to cutting-edge machine learning methodologies. We encourage astronomers to embrace these advancements, fostering a more streamlined and accurate segmentation process that aligns with the ever-expanding frontiers of astronomical exploration.

\end{abstract}



\end{frontmatter}


\section{Introduction}
\label{Introduction}
Astronomy is currently experiencing a data-driven revolution, fueled by an exponential increase in data acquisition. Numerous comprehensive sky surveys have already been completed, with many more planned \citep{2015JATIS...1a4003R,2019PASP..131a8002B,2019ApJ...873..111I,2019AJ....157..168D,2020ApJS..251....7F,2020ApJS..249....3A,2021A&A...649A...1G}. This deluge of data poses significant challenges to astronomers, particularly in the complex analysis of these voluminous datasets. A fundamental and recurring task in astronomy is segmentation, the process of accurately identifying celestial objects or features \citep{1996A&AS..117..393B,2015A&C....10...22B,2019ascl.soft07016R}.

Image segmentation is an essential first step in analyzing images or three-dimensional data cubes. Its primary goal is to partition this data into distinct segments or regions, enabling the differentiation of boundaries between individual celestial objects. This results in a robust foundation for further scrutiny, enabling a comprehensive examination of the unique attributes associated with each distinct component. Astronomical images and data cubes inherently encompass a diverse array of celestial objects and phenomena. Segmentation is instrumental in separating these multifarious elements, enabling an in-depth examination of individual components.


The ability to scrutinize the dimensions, contours, and assorted characteristics of these segmented regions equips astronomers with a heightened level of precision in their research pursuits and analytical comparisons. For example, segmentation of galaxies delineates their inherent components—such as spiral arms, bulges, and bars—thereby elucidating the complex processes underlying their formation and evolution \citep{2020RAA....20..159S,2021A&A...647A.120B,2023BAAA...64..253R}.

Furthermore, the application of segmentation extends to the analysis of multi-wavelength data, contributing to a holistic understanding of targeted astronomical subjects. This approach involves exploring a broad spectrum of wavelengths, ranging from radio waves and infrared to optical, UV, x-ray, and gamma-rays \citep{2012A&A...542A..81M,2013SoPh..283...67V,2019A&A...628A..33R,2021SoPh..296..138V,2021SoPh..296...71A,2023PASA...40....1H}. Additionally, segmentation is employed in the analysis of gravitational wave data \citep{PhysRevD.105.124007} and neutrino observations \citep{Belavin_2021}. The diverse range of wavelengths and messengers provides distinct insights into various physical processes, elevating the depth of the comprehensive analysis of celestial objects.


Additionally, segmentation facilitates the study of transient or time-dependent astronomical phenomena, a domain that encompasses celestial events like supernovae and variable stars \citep{2008SoPh..248..485O,2014A&A...561A..29V}. In this context, segmentation becomes the conduit through which astronomers can trace the temporal evolution and transformations of these celestial bodies, thereby facilitating comprehensive investigations into the nature of these transient occurrences.

As the volume of observational data swells, the importance of automation through segmentation is underscored. Leveraging the capabilities of machine learning and deep learning techniques, astronomers can effectively automate the process of identifying and categorizing objects within expansive datasets, thus enhancing the efficiency and efficacy of astronomical analyses.

In this review paper, we embark on a two-fold journey: an exploration of classical segmentation methods in Section \ref{Classic Segmentation Methods}, followed by an investigation into machine learning-based approaches to segmentation in Section \ref{Segmentation Based on Deep Learning}.

\section{Classic Segmentation Methods}
\label{Classic Segmentation Methods}

In this section, we introduce several classical segmentation method in astronomy. 


\subsection{Thresholding}
\label{Thresholding}

Thresholding is a fundamental and well-established image segmentation technique, widely used in image analysis and processing, holding particular significance in the realm of astronomy. It plays a pivotal role in partitioning astronomical images into discrete regions or objects by distinguishing between pixel values above and below a predetermined threshold. Thresholding's essence lies in its elementary and unpretentious nature, serving as the gatekeeper that demarcates an image into two or more segments based on the intrinsic pixel intensity values.

Determining a suitable threshold poses challenges, given factors like noise, background variations, or the diffuse boundaries of objects. Selecting a threshold involves a trade-off: minimizing false negatives may lead to an increase in false positives, and vice versa. Striking a balance between these errors becomes challenging, as extreme threshold variations tend to minimize one error type at the expense of the other. Consequently, the task lies in finding a threshold that minimizes both types of errors simultaneously.

Various strategies exist for setting the threshold, often adopting arbitrary methods. Examples include adjusting it based on sky background and noise levels \citep[e.g., ]{1985MNRAS.214..575I,1999A&AS..138..365S} or deriving it from modeled distributions, such as $\chi^{2}$ distributions. In the latter case, the threshold is determined at the intersection point between the theoretical and actual data distributions \citep{1999AJ....117...68S}. Alternatively, if the emission distribution of real sources is known, one might opt for a threshold set at three times the deviation of the peak distribution \citep{1988A&A...201....9S} or choose a threshold that minimizes the fraction of false detections \citep{1999ApJ...524..414L,2002AJ....123.1086H}. However, it is crucial to note that these methods are not fully automated, and the selection of the threshold may involve some level of arbitrariness. 

The majority of source extraction algorithms in astronomy, such as PyBDSF \citep[Python Blob Detection and Source Finder,][]{2015ascl.soft02007M}, Selavy \citep{2012PASA...29..371W}, Aegean \citep{2012MNRAS.422.1812H}, and Caesar \citep[Compact And Extend Source Automated Recognition,][]{2018ascl.soft07015R}, rely on basic thresholding as their fundamental approach, with some variation in the treatment of threshold selection among different algorithms. \xd{Additional algorithms employing direct thresholding on optical, x-ray, and multiband images have been developed, as demonstrated in studies by \citet{1977ApJS...33...55H,1977PASP...89..925N,1983A&A...126..278B,1995ApJ...451..542V}.}

One noteworthy instance within the realm of thresholding is Otsu's method, proposed by Nobuyuki Otsu in 1979 \citep{otsu1979threshold}. This method is exceptionally proficient in the automated selection of an optimal threshold for image segmentation. Otsu's method optimizes the between-class variance of pixel intensities, with the primary objective of demarcating the object of interest from the background. Notably, Otsu's method excels when confronted with an image whose pixel intensity histogram exhibits a bimodal or multimodal distribution, a common scenario in image segmentation.

The Otsu thresholding algorithm can be summarized in the following steps:

\begin{enumerate}
\item Histogram computation: Compute the histogram of the input image, revealing the distribution of pixel intensities.

\item Probability distribution: Normalize the histogram to obtain a probability distribution of pixel intensities.

\item Threshold initialization: Initialize a threshold value, typically set to zero.

\item Between-class variance calculation: For each feasible threshold value, calculate the between-class variance, which quantifies the dispersion of pixel intensities between the object of interest and the background.

\item Within-class variance calculation: Concurrently, for each threshold value, calculate the within-class variance for both the background and the object of interest, which characterizes the spread of pixel intensities within these classes.

\item Weighted sum of variances: Compute a weighted sum of variances for each threshold. This measure captures the amalgamation of the between-class and within-class variances.

\item Threshold selection: Iterate through steps 4-6, identifying the threshold value that optimally maximizes the total variance. This optimal threshold signifies the zenith of contrast and separation between the object of interest and the background in the image.

\item Binarization: Apply the selected threshold to the original image, rendering a binary image in which the object of interest and the background are distinctly differentiated, setting the stage for further image processing or analysis.

\end{enumerate}

\begin{figure}[!htbp]
    \centering
    \includegraphics[width=0.9\linewidth]{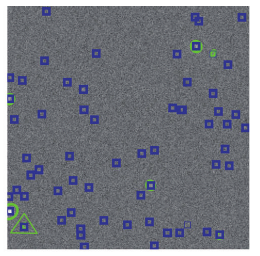}
    \caption{Captured from \citet{2015MNRAS.451.4445Z}, squares indicate accurately known object positions. Ellipses represent objects identified using the enhanced Otsu's method outlined in \citet{2015MNRAS.451.4445Z}, while triangles denote true, faint objects exclusively detected by this method but not by SExtractor.}
    \label{fig_zhnag15_otsu}
\end{figure}

Otsu's method is a robust and versatile thresholding algorithm that has been adopted in a diverse range of scientific and engineering fields, such as astronomical image segmentation for the identification of stars and galaxies \citep{2015MNRAS.451.4445Z}. Its simplicity and efficiency render it a valuable tool for image segmentation and other image processing tasks. It is noteworthy that SExtractor \citep{1996A&AS..117..393B}, a extensively utilized algorithm for source segmentation, also employs the thresholding concept during the process of source identification. Illustrated in Figure~\ref{fig_zhnag15_otsu} is an instance of source segmentation/detection utilizing the improved Otsu's method introduced by \citet{2015MNRAS.451.4445Z}.


\subsection{Edge Detection}
\label{Edge Detection}

Edge detection is a fundamental image processing technique that identifies discontinuities in pixel intensity values, corresponding to the boundaries of objects and regions. It plays a pivotal role in image segmentation, providing the foundation for subsequent analysis.

In the realm of astronomy, where precision and accuracy are paramount, various edge detection algorithms are applied, each offering distinct advantages and trade-offs. Notably, the Sobel, Prewitt, and Canny methods are three prominent gradient-based techniques widely adopted for astronomical image analysis.

The Sobel and Prewitt operators are gradient-based edge detection algorithms that operate by convolving the image with a pair of $3\times3$ kernels, one for horizontal edges and one for vertical edges. The Sobel and Prewitt kernels are as follows:

Sobel: $G_{x}= \begin{bmatrix}
+1 & 0 & -1\\
+2 & 0 & -2 \\
+1 & 0 & -1\\
\end{bmatrix}
$, $G_{y}= \begin{bmatrix}
+1 & +2 & +1\\
0 & 0 & 0 \\
-1 & -2 & -1\\
\end{bmatrix}
$,

Prewitt: $G_{x}= \begin{bmatrix}
+1 & 0 & -1\\
+1 & 0 & -1 \\
+1 & 0 & -1\\
\end{bmatrix}
$, $G_{y}= \begin{bmatrix}
+1 & +1 & +1\\
0 & 0 & 0 \\
-1 & -1 & -1\\
\end{bmatrix} 
$. 

The convolution operation computes the weighted sum of the neighboring pixels, where the weights are given by the kernel. The resulting images, $G_{x}$ and $G_{y}$, represent the horizontal and vertical gradients of the input image, respectively. The edge magnitude and direction can then be calculated from the gradient images:
\begin{align*}
{\rm magnitude}&=\sqrt{G_{x}^2 + G_{y}^2}\\
{\rm direction}&=atan2(G_{y},(G_{x}).
\end{align*}
Pixels with a high magnitude and a consistent direction are considered to be edges. The Sobel and Prewitt operators are simple and efficient, but they can be sensitive to noise in the image. To improve robustness to noise, it is common to pre-smooth the image with a Gaussian filter before applying the edge detection algorithm.

In contrast to the Sobel and Prewitt operators, the Canny edge detector distinguishes itself as a multi-stage edge detection approach celebrated for its robustness and precision. The Canny method incorporates a sequence of stages:
\begin{enumerate}
\item Gaussian smoothing: The image is smoothed with a Gaussian filter to reduce noise and create a continuous gradient.

\item Gradient calculation: The gradient magnitude and direction are calculated at each pixel location.

\item Non-maximum suppression: Only the local maxima along the gradient direction are preserved, resulting in single-pixel-wide edges.

\item Edge tracking by hysteresis: A two-level thresholding strategy is used to link strong edges together and discard weak edges.

\end{enumerate}

The Canny edge detector is more computationally expensive than other edge detection algorithms, such as the Sobel and Prewitt operators, but it produces more accurate edge maps. It is also more robust to noise, making it suitable for a wide range of applications in computer vision, image processing, and image analysis. 

Edge detection is a fundamental image processing technique that is widely used in astronomy to identify structures. For example, the Sobel operator is used to detect filamentary structures in molecular clouds, which can be used to study their alignment with the magnetic field direction \citep{2017ApJS..232....6G}. The Sobel edge detection technique is also used to measure the magnitude of the tip of the red giant branch (TRGB), which is a key parameter for determining distances to nearby galaxies \citep{2005ApJ...633..810M}. Additionally, the Canny edge detector is used to detect transient events such as coronal mass ejections from the Sun \citep{2005SPIE.5901...13B}, and to search for the signatures of cosmic string networks on cosmic microwave background (CMB) anisotropies \citep{2018MNRAS.475.1010V}.

\begin{figure}[!htbp]
    \centering
    \includegraphics[width=0.99\linewidth]{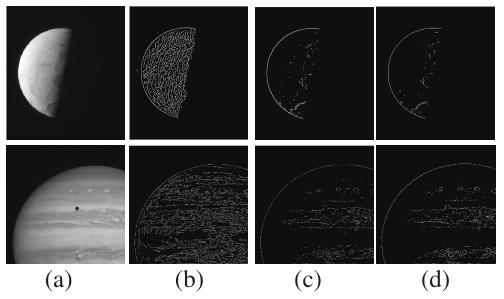}
    \caption{Detected edges using various methods on Cassini Astronomy images, as detailed in \citet{yang2018edge}. (a) Depicts the Original Image. (b), (c), and (d) illustrate the edges detected by Canny, Sobel, and Prewitt methods, respectively.}
    \label{fig_Yang_edge_all}
\end{figure}

The Sobel, Prewitt, and Canny methods are all fundamental edge detection techniques, each with its own strengths and weaknesses. The choice of which method to use depends on factors such as the level of image noise, the required precision, and the computational constraints. Figure~\ref{fig_Yang_edge_all} illustrates the detected edges in Cassini Astronomy images employing different methods such as Canny, Sobel, and Prewitt \citep{yang2018edge}.

\subsection{Watershed}

Watershed segmentation is a widely used image segmentation algorithm that partitions an image into distinct regions based on pixel intensity and spatial relationships. The algorithm works by treating the image as a topographic surface, with pixel intensity values representing elevations. Water is flooded onto the surface from markers, which are typically identified by the user or generated using image processing techniques. The water flows into valleys and basins, and the watershed lines (i.e., the boundaries between the basins) define the segmented regions. Here is a step-by-step overview of the watershed segmentation process:
\begin{enumerate}
\item Preprocessing: Denoise and enhance the image, if needed.

\item Gradient computation: Calculate the image gradient magnitude to emphasize areas of rapid change in intensity.

\item Marker selection: Identify seed pixels, either manually or using image processing techniques.

\item Distance transform: Calculate the distance transform of the markers to define their potential regions of influence.

\item Watershed flooding: Simulate water rising from the markers and flooding the image, separating basins with watershed lines.

\item Region merging: Merge adjacent basins as the water level rises.

\item Result visualization: Label each pixel in the segmented image according to its basin.

\item Post-processing: Refine the segmentation results as needed, e.g., by removing small regions or merging adjacent regions.
\end{enumerate}
Watershed segmentation is a powerful image segmentation algorithm, but it is sensitive to the selection of markers and initial conditions. Careful marker selection and post-processing are essential for accurate and meaningful results.

\begin{figure}[!htbp]
    \centering
    \includegraphics[width=0.99\linewidth]{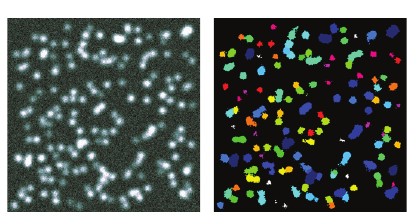}
    \caption{From \citet{2015A&C....10...22B}, on the left: an artificial field containing Gaussian clumps. On the right: the array depicting the clump assignments generated by FellWalker.}
    \label{fig_fellwalker_2015}
\end{figure}

The watershed algorithm is widely used in astronomy to segment a variety of objects, including filamentary structures in molecular clouds \citep[e.g., FELLWALKER method, ][]{2015A&C....10...22B,2023MNRAS.523.1832R}, stars, galaxies \citep{2015MNRAS.451.4445Z,2020ApJS..248...20H}, and large-scale structures such as voids \citep{2007MNRAS.380..551P}. It is worth mentioning that the general source extraction code ProFound \citep{2018MNRAS.476.3137R,2019MNRAS.487.3971H} also incorporates the watershed algorithm. Figure~\ref{fig_fellwalker_2015} illustrates an instance of Gaussian clump segmentation using FellWalker \citep{2015A&C....10...22B}.

\subsection{Active Contours}

Active contours, also known as snakes, are a powerful image segmentation technique that uses an iterative optimization algorithm to minimize an energy functional. The energy functional consists of two terms: an internal energy term that penalizes the curvature and elasticity of the contour, and an external energy term that attracts the contour to image features such as edges and lines. Active contour segmentation is typically performed as follows:
\begin{enumerate}
\item Initialize the active contour, a process that can be performed either manually by the user or automatically through an image processing algorithm. One straightforward approach is to apply image thresholding. 
\item Compute the energy functional. The energy functional is computed based on the current position of the active contour and the image features.
\item Deform the active contour. The active contour is deformed to minimize the energy functional. This can be done using a variety of optimization algorithms.
\item Repeat steps 2 and 3 until the active contour converges. Convergence occurs when the energy functional cannot be further reduced.
\item The final position of the active contour represents the segmented object.
\end{enumerate}

\begin{figure}[!htbp]
    \centering
    \includegraphics[width=0.9\linewidth]{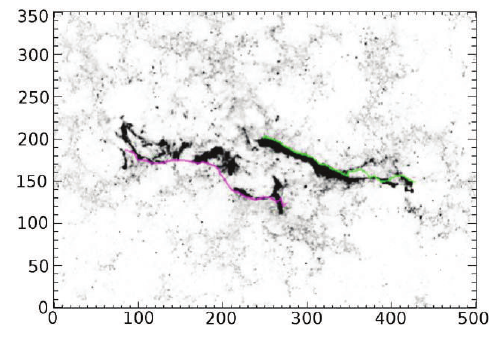}
    \caption{Tracking ribbon positions using active contours on a solar-flare UV and EUV image from \citet{2010SoPh..262..355G}. }
    \label{fig_activa_contour_UV}
\end{figure}

Active contour segmentation is a powerful and versatile technique for segmenting objects with a wide range of shapes and appearances. However, it is important to note that the accuracy of the segmentation results is sensitive to the choice of initial contour and the parameters of the energy functional. Active contour segmentation is employed in solar astrophysics for various applications.  Its usage spans the segmentation of elongated bright ribbons in solar-flare UV and EUV images, contributing to the investigation of magnetic field line reconnection \citep{2010SoPh..262..355G}. Furthermore, active contour segmentation is employed to track and determine the differential rotation of sunspots and coronal bright points \citep{2014SunGe...9...81D}. In the detection of coronal holes, it integrates a confidence map for enhanced accuracy \citep{grajeda2023quantifying}. Additionally, active contour segmentation facilitates real-time detection and extraction of coronal holes from solar disk images \citep{Bandyopadhyay2020}. It is also utilized to segment diffuse objects within coronal holes \citep{tajik2013diffuse} and extract as well as characterize coronal holes, playing a crucial role in predicting the fast solar wind \citep{2016SoPh..291.2353B}. 
Figure~\ref{fig_activa_contour_UV} depicts the utilization of the active contour method for tracking ribbon positions on a solar-flare UV and EUV image \citep{2010SoPh..262..355G}.

\subsection{Wavelet Transform}
\label{Wavelet Transform}

Wavelet transform, originally designed for signal processing, has become a versatile tool in image processing and segmentation. It excels at decomposing images into frequency components, enabling simultaneous analysis in spatial and frequency domains. In image segmentation, wavelet transform represents features at different scales through multiresolution analysis. Commonly, wavelet-based segmentation uses adaptive thresholding to separate regions of interest from background noise. This adaptability ensures nuanced segmentation, making wavelets robust for diverse image structures, including those encountered in medical imaging, remote sensing, and astronomy.

The ``à trous" algorithm, also known as the "with holes" algorithm or Stationary Wavelet Transform (SWT), is a pivotal wavelet-based method in image processing and segmentation. It facilitates multiresolution analysis by decomposing an image into frequency components at various scales without explicit downsampling, preserving the original sampling grid and leaving ``holes" in the transformation process. This unique approach allows for a detailed analysis of images at different scales without loss of information. In image segmentation, the algorithm accurately represents image features, capturing both global and local characteristics. Its adaptability makes it well-suited for diverse segmentation tasks, including those in astronomy with varying intensities in astronomical images. The ``à trous" algorithm's distinctive approach enhances the precision of image analysis, contributing to the broader field of image processing and analysis. \xd{We utilize the mathematical framework described in \citet{starck1998image,1999A&AS..138..365S}. The ``à trous" algorithm decomposes an image into different levels (typically $J$ levels). For each decomposition level ($j$ = 0 to $J$-1), the process involves convolving the original image ($f(x,y)$) with a scaling function ($h$) to obtain smoothed data ($c_j(k,l)$) at resolution level ($j$) and position ($k$,$l$). This is mathematically represented as:
\begin{align*}
{c_{0}(k,l)}&=f(k,l)\\
c_j(k,l) &= \sum_{m,n} h(m,n) \cdot c_{j-1}(k + 2^{j-1}m, l + 2^{j-1}n).
\end{align*}
Here, ($m$,$n$) ranges over the filter size. The difference signal ($w_j$) between two consecutive resolutions is computed as:
\begin{align*}
w_j(k,l) = c_{j-1}(k,l) - c_j(k,l).
\end{align*}}




This algorithm constructs the sequence by performing successive convolutions with a filter derived from an auxiliary function known as the scaling function $h$. \xd{A commonly employed linear profile scaling function
$h$ is typically represented as follows:
\begin{align*}
h&=\frac{1}{16}\begin{pmatrix}
1 & 2 & 1 \\
2 & 4 & 2 \\
1 & 2 & 1\\
\end{pmatrix}.
\end{align*}}
A widely used B3 cubic spline profile scaling function $h$ takes the following form:
\begin{align*}
h&=\frac{1}{256}\begin{pmatrix}
1 & 4 & 6 & 4 & 1\\
4 & 16 & 24 & 16 & 4\\
6 & 24 & 36 & 24 & 6\\
4 & 16 & 24 & 16 & 4\\
1 & 4 & 6 & 4 & 1\\
\end{pmatrix}.
\end{align*}

\begin{figure}[!htbp]
\centering
\includegraphics[width=0.99\linewidth]{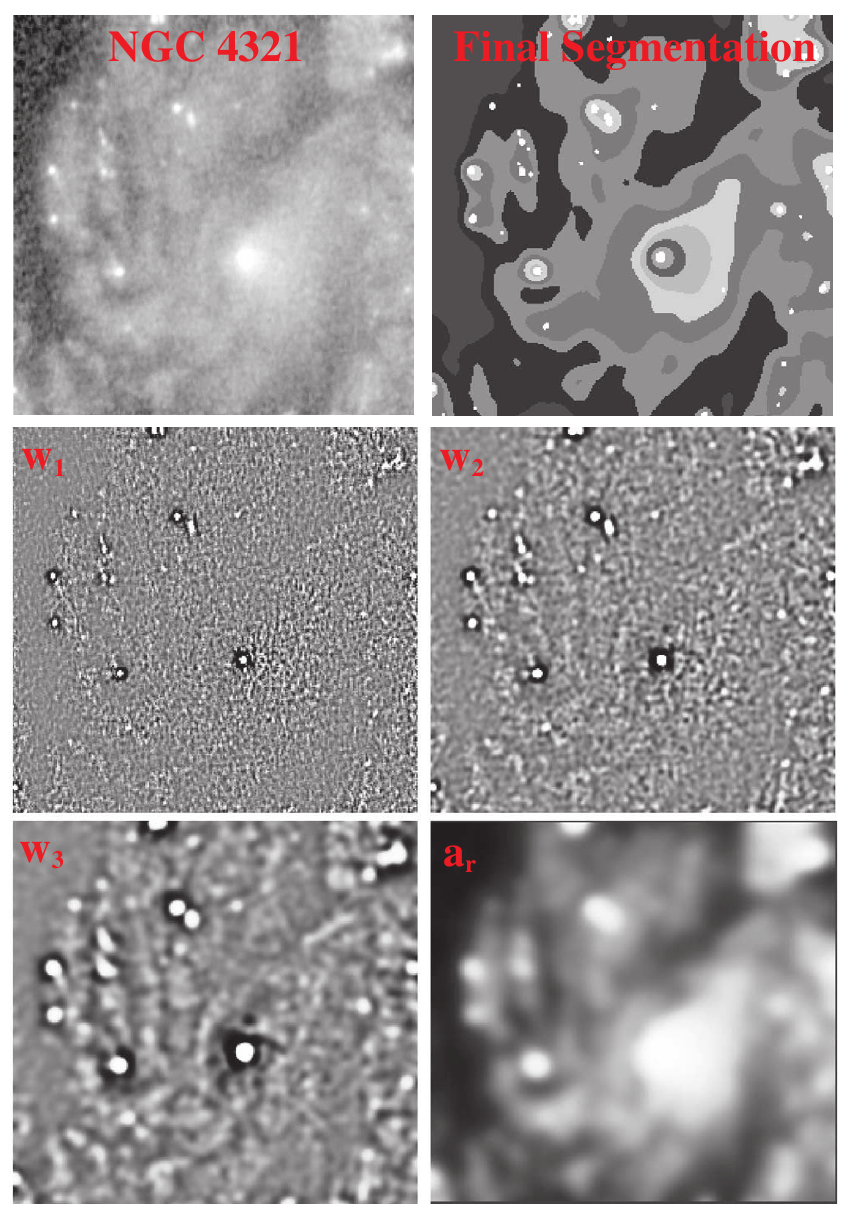}
\caption{NGC 4321 image (upper left), segmentation of the NGC 4321 image (upper right), wavelet decomposition planes $W_{1}$ (middle left), $W_{2}$ (middle right), $W_{3}$ (lower left), and residual image $a_{r}$ (lower right), as presented in \citet{nunez2003astronomical}.}
\label{fig_wavelet_ngc4321}
\end{figure}

Below are the steps involved in using the ``à trous" algorithm for image segmentation:
\begin{enumerate}
\item Wavelet Decomposition: Apply the "à trous" algorithm to decompose the input astronomical image into frequency components at multiple scales through multiresolution analysis. 

\item Wavelet Coefficient Thresholding: Threshold the obtained wavelet coefficients at different scales to distinguish significant features from background noise. Adaptive thresholding ensures nuanced segmentation, accommodating varying complexities in different astronomical structures, including objects with diverse intensities.

\item Segmentation: Utilize the thresholded wavelet coefficients to perform precise segmentation on the astronomical image. The algorithm's capability to capture both global and local features makes it well-suited for accurately representing complex structures during the segmentation process in astronomical data.

\end{enumerate}

The ``à trous" algorithm excels in multiresolution analysis, decomposing images into frequency components at various scales to capture information at different levels of detail. Unlike traditional wavelet methods, it avoids explicit downsampling, preserving the original sampling grid for a detailed analysis of images without information loss. Adaptive thresholding of wavelet coefficients enables nuanced segmentation, adapting to complexities in different image structures, particularly beneficial for astronomical images with varying intensities. The algorithm accurately represents global and local characteristics, making it suitable for diverse segmentation tasks. It demonstrates robustness in handling irregularities, noise, and contrast variations. However, computational intensity arises with large datasets or images, escalating with the number of scales in multiresolution analysis. Sensitivity to parameter choices, such as scaling function and scale count, may necessitate experimentation for optimal performance. Challenges exist in precisely localizing features, especially when boundaries are ambiguous. The choice of scaling function influences performance, with certain functions better suited for specific image types or structures.

In the field of astronomy, the "à trous" algorithm has found application in image segmentation \xd{\citep{rue1996pyramidal,barra2009fast, Xavier2012AnEA, chen2023fast,ellien2021dawis}}, particularly in tasks such as segmenting various galaxy components \citep{BIJAOUI1995345, nunez2003astronomical}. Figure~\ref{fig_wavelet_ngc4321} provides an illustrative example of the "à trous" algorithm's effectiveness in segmenting different components of NGC 4321. Moreover, the algorithm has been widely employed in segmenting extended structures, including supernova remnants, HII regions, and bow shocks in radio images \citep{IEEE6116254}.

\subsection{Clustering-based Methods}
\label{Clustering-based Methods}

\begin{figure}[!htbp]
    \centering
    \includegraphics[width=0.99\linewidth]{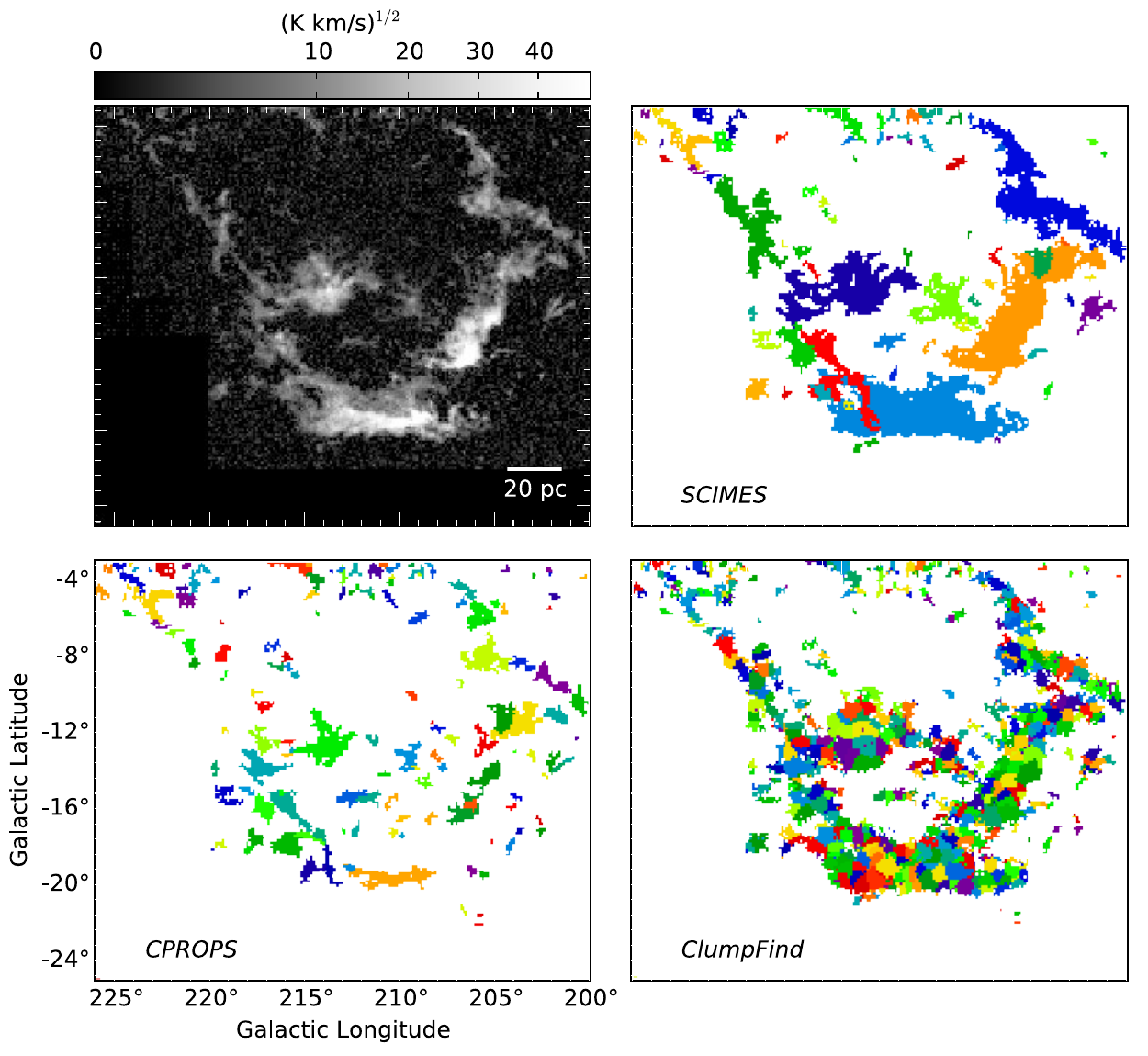}
    \caption{From \citet{2015MNRAS.454.2067C}, top left: the Orion–Monoceros complex. Top right: SCIMES cloud decomposition using spectral clustering on the dendrogram of emission \citep{2015MNRAS.454.2067C}. Bottom left: structures identified by CPROPS' island method based on dendrogram \citep{2006PASP..118..590R}. Bottom right: CLUMPFIND cloud decomposition using a "friends-of-friends" algorithm \citep{1994ApJ...428..693W}. }
    \label{fig_Colombo_15_mn_crop}
\end{figure}

Clustering-based methods are a widely adopted and versatile approach to image segmentation, and their significance extends into the domain of astronomy. These methods are prized for their versatility, serving as foundational tools for partitioning astronomical images into meaningful regions and objects. The core premise revolves around the grouping of pixels with akin characteristics into clusters, thereby facilitating the creation of distinctive image segments. This process hinges on the underlying principle that pixels sharing common attributes, be it intensity values or color characteristics, exhibit greater similarity within their designated cluster than to pixels belonging to other clusters. Notable clustering techniques deployed in this pursuit encompass:
\begin{enumerate}
\item[\large\textbullet] K-Means clustering: K-Means clustering is a simple and efficient clustering algorithm that is widely used in image segmentation. K-Means clustering works by first partitioning the image pixels into a predefined number of clusters. Then, each pixel is assigned to the cluster with the nearest centroid. The centroids are then updated to reflect the new cluster assignments. This process is repeated until no further changes in the cluster assignments occur. K-Means clustering is a relatively simple algorithm to implement and is computationally efficient, but it can be sensitive to the choice of the initial centroids and the number of clusters.

\item[\large\textbullet] Fuzzy C-Means (FCM) clustering: FCM clustering is a more flexible clustering algorithm that allows pixels to belong to multiple clusters with different degrees of membership. This flexibility allows FCM clustering to produce more nuanced and accurate segmentation results than K-Means clustering, especially for images with complex or overlapping objects. However, FCM clustering is computationally more expensive than K-Means clustering.

\item[\large\textbullet] Density-Based Spatial Clustering of Applications with Noise (DBSCAN): DBSCAN is a clustering algorithm that is robust to noise and outliers. DBSCAN works by grouping pixels based on their density and spatial proximity. A pixel is considered to be a core point if it has a minimum number of neighboring pixels within a specified distance threshold. Core points are then assigned to clusters, and the clusters are expanded to include all neighboring pixels within the distance threshold. This process continues until all pixels have been assigned to a cluster or marked as noise.

\item[\large\textbullet] Hierarchical clustering:  Hierarchical clustering is a clustering algorithm that organizes pixels into a hierarchical tree structure. Hierarchical clustering works by iteratively merging or splitting clusters based on the similarity of their constituent pixels. The termination condition for hierarchical clustering can be a predefined number of clusters or a specific stopping criterion. Hierarchical clustering allows for multi-level analysis of complex images, but it can be computationally expensive for large images.
\end{enumerate}

Clustering-based methodologies play a pivotal role in image segmentation, facilitating the meticulous identification and delineation of regions and objects in images. These techniques are employed in astronomy for diverse segmentation tasks. For example, FCM clustering has been instrumental in partitioning EUV solar images into well-defined regions, such as active regions, coronal holes, and the quiet sun \citep{2009A&A...505..361B,2014A&A...561A..29V}. Additionally, DBSCAN has proven highly effective in segmenting molecular cloud emissions while robustly mitigating noise contamination in astronomical data \citep{2020ApJ...898...80Y}. Notably, \citet{2014A&A...568A..56J} skillfully harnessed hierarchical clustering, exemplified through the implementation of Astrodendro \citep{2019ascl.soft07016R}, to segment molecular cloud emissions in astronomical investigations. Figure~\ref{fig_Colombo_15_mn_crop} illustrates the results of emission segmentation using various algorithms on the Orion–Monoceros complex.

\subsection{Multiband Segmentation}

It is noteworthy that several classical segmentation methods mentioned earlier are predominantly designed for single-band images rather than multiband ones. Although one can apply these methods individually to each single-band image and subsequently merge the results based on specific criteria for a comprehensive multiband segmentation, there are techniques specifically crafted to handle multiband images. Examples of such methods include the Hierarchical Hidden Markov Model (HHMM) \citep{COLLET20042337} and the Connected Component Tree (cc-trees) \citep{Slezak2010cctree}, both falling under the hierarchical clustering category discussed earlier.

\begin{figure}[!htbp]
    \centering
    \includegraphics[width=0.99\linewidth]{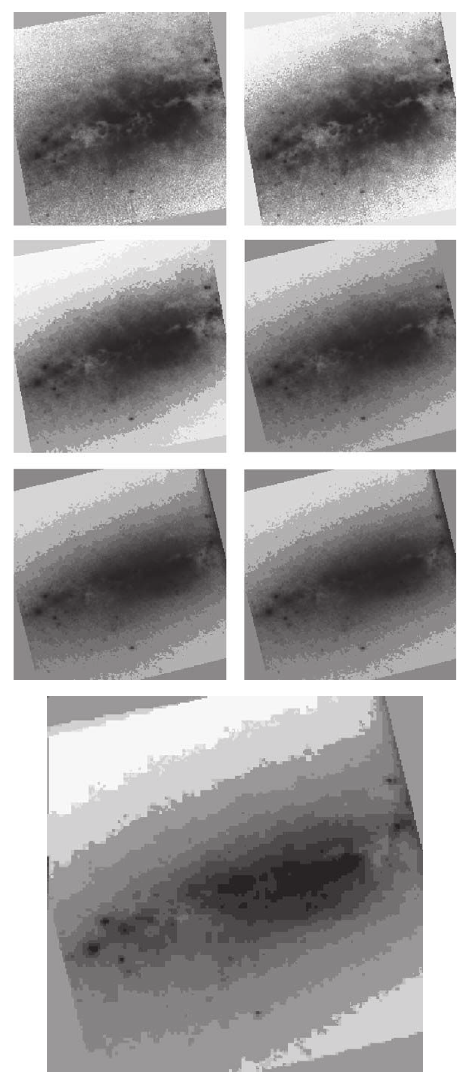}
    \caption{From \citet{COLLET20042337}, presentation of 6-spectral band images depicting the M82 starburst galaxy (rows 1-3), alongside the HHMMS segmentation results applied to the astronomical 6-spectral band images from the M82 region (row 4). }
    \label{fig_HHMM_galaxy}
\end{figure}

The Hierarchical Hidden Markov Model (HHMM) serves as a potent Bayesian estimation framework for image segmentation, offering probabilistic sequences of observations. Its distinguishing characteristic lies in the "hidden" nature of the underlying stochastic process generating observations, governed by the Markov property wherein the state at any time depends solely on the preceding state. Through its hierarchical structure, the HHMM adeptly captures intricate dependencies and patterns in data, making it ideal for image segmentation tasks. This model excels in unraveling complex visual structures, leveraging Markovian principles across various image levels to discern dependencies and relationships. By modeling both global and local information, HHMMs effectively handle complex image structures, with transitions between states enhancing segmentation precision. The hierarchical design allows for adaptability in revealing varying detail levels. Notably, computational demands for HHMM training and inference, especially with large images, can be substantial. Additionally, optimizing the hierarchical structure and hidden state count poses challenges. Figure~\ref{fig_HHMM_galaxy} illustrates HHMM application in segmenting galaxy components \citep{COLLET20042337}.

\begin{figure}[!htbp]
    \centering
    \includegraphics[width=0.99\linewidth]{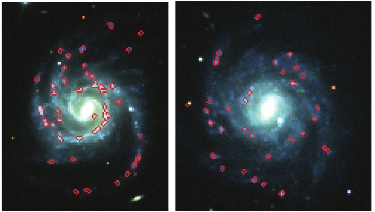}
    \caption{From \citep{Slezak2010cctree}, detection of HII regions using cc-trees in two 5-band astronomical images.}
    \label{fig_cctree_hii}
\end{figure}

The Connected Component Tree (cc-tree) is a hierarchical data structure employed in image segmentation for delineating connected components or regions within an image. Each node within the cc-tree corresponds to a distinct connected component, capturing the relationships and hierarchy among these components. The construction process begins with identifying connected components based on criteria such as pixel intensity or color similarity, followed by iterative hierarchical merging or splitting. The resulting cc-tree provides a comprehensive representation of the image's structure, aiding the understanding of spatial organization and relationships among different regions. CC-Trees are widely used in image segmentation tasks, especially in scenarios with hierarchical or detailed object organization. They offer a versatile tool for applications like object recognition and feature extraction, efficiently representing segmented objects and their relationships. Figure~\ref{fig_cctree_hii} exemplifies the application of cc-trees in segmenting HII regions in a 5-band galaxy observation \citep{Slezak2010cctree}. However, the efficacy of cc-trees depends on the quality of the initial image segmentation; errors may occur if segmentation leads to over-segmentation or smaller, disconnected components, affecting the accuracy of the cc-trees in capturing true object relationships.

It is important to highlight that there is no distinction in deep learning approaches between single-band segmentation, multiband segmentation, or even 3D data cube segmentation. This lack of difference arises from the flexibility of input data during deep learning model training, a topic that will be explored in the subsequent section.

\section{Segmentation Based on Deep Learning}
\label{Segmentation Based on Deep Learning}

In this section, we introduce various deep learning approaches to image segmentation in astronomy. Given the rapid evolution of the deep learning field, some state-of-the-art algorithms have limited applications or are yet to be explored in astronomy. The purpose of this section is to draw attention to applicable deep learning methods, informing astronomers about efficient and high-performance approaches that can enhance their tasks.

The general steps involved in training and applying deep learning methods for astronomy segmentation tasks include the following:
\begin{enumerate}
\item Data Preparation: Assemble a diverse dataset with labeled images for training and validation. Clean and preprocess the data by resizing, normalizing, and augmenting to improve model generalization.

\item Model Selection: Select an appropriate deep learning architecture (e.g., U-Net, Mask R-CNN) based on task requirements and available resources. Choose between building a custom structure or adopting pretrained models for transfer learning.

\item Model Configuration: Fine-tune parameters like learning rate, batch size, and regularization for optimal performance. Define a suitable loss function aligned with segmentation objectives (e.g., cross-entropy loss, Dice loss, as discussed in Section~\ref{Metrics and Evaluations}).

\item Training: Feed input images into the network and optimize the model to minimize the defined loss function. Monitor performance on a validation set to prevent overfitting.

\item Evaluation: Assess the model's performance using evaluation metrics such as IoU, Dice coefficient, or pixel-wise accuracy.

\item Fine-Tuning (Optional): If necessary, iteratively fine-tune the model based on evaluation results.

\item Inference: Evaluate the model on an independent testing set to measure real-world performance. Apply the trained model to segment objects in new, unseen images.

\item Post-Processing (Optional): Implement post-processing techniques to improve segmentation results and address artifacts.

\end{enumerate}

These steps offer a general framework, and specific details may vary based on the chosen model architecture, dataset characteristics, and segmentation task requirements. Further discussion on these aspects will be provided in the subsequent section.






\subsection{Mask R-CNN (Region-Based Convolutional Neural Network)}

\begin{figure}[!htbp]
    \centering
    \includegraphics[width=0.48\textwidth]{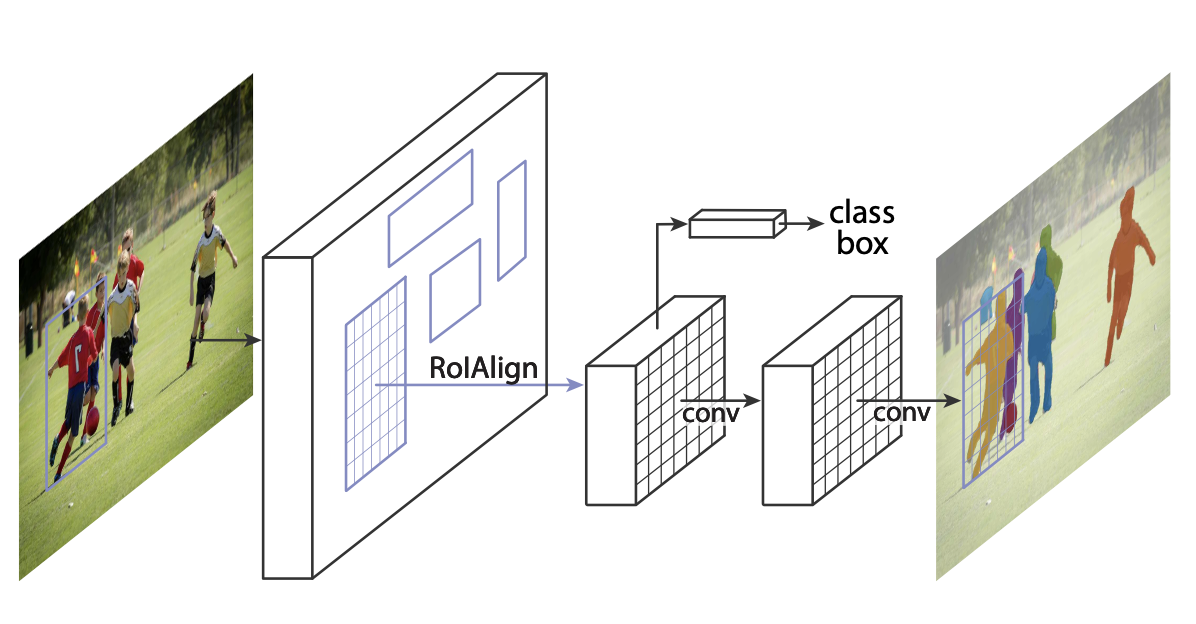}
    \caption{Overview of the Mask R-CNN framework, as presented in~\cite{he2017mask}.}
\label{fig:mask-rcnn}
\end{figure}

Mask R-CNN~\citep{he2017mask}, a deep learning extension of Faster R-CNN~\citep{girshick2015fast}, has revolutionized image segmentation, particularly in astronomy. The precise delineation of celestial objects in astronomical images is crucial, and Mask R-CNN excels in this domain by enabling pixel-level object segmentation. Mask R-CNN is a region-based convolutional neural network that is finely tuned for detecting and segmenting objects in images. Building on the success of Faster R-CNN, a leading object detection model, Mask R-CNN has become instrumental in image segmentation tasks.

Figure~\ref{fig:mask-rcnn}, taken from the work by \citet{he2017mask}, illustrates the workflow of Mask R-CNN. In addition to the general steps introduced at the beginning of this section, we elaborate on the essential elements and central processes of Mask R-CNN as follows:
\begin{enumerate}
\item Region Proposal Network (RPN): The RPN takes the extracted features from the backbone network and generates a set of proposal boxes, each associated with a score indicating the likelihood of an object being present within the box.

\item Region of Interest (RoI) Align: The RoI Align operation takes the proposal boxes and extracts corresponding feature maps from the backbone network, preserving spatial information and aligning the feature maps with the regions.

\item Classification and Bounding Box Refinement: For each proposal box, Mask R-CNN performs two tasks: classification and bounding box refinement. It first classifies the object within the box and refines the box's position if necessary.

\item Mask Generation: Mask R-CNN generates a segmentation mask for each object within the proposal boxes using a separate neural network that predicts pixel-wise masks for each object.

\item Final Prediction: The final output of Mask R-CNN includes: 
\begin{itemize}
  \item Classification results indicating the object class for each box
  \item Refined bounding box coordinates
  \item Segmentation masks representing the object's precise shape at the pixel level
\end{itemize}

\item Post-Processing: The final results can undergo post-processing steps to filter out low-confidence detections, eliminate duplicate detections, and fine-tune the masks.

\item Output: The output of the Mask R-CNN algorithm is a set of bounding boxes, each associated with a class label and a high-resolution binary mask that accurately delineates the object.

\end{enumerate}

Mask R-CNN demonstrates resilience to noise and has maintained a track record of excellence in numerous image segmentation challenges and competitions. However, it is important to note that the computational demands of training and deploying a Mask R-CNN model can be substantial, necessitating the availability of robust GPUs. Additionally, this model often mandates a substantial volume of meticulously annotated data for effective training, a process that can be labor-intensive and costly. In certain scenarios, Mask R-CNN has the potential to yield an abundance of smaller segments, contributing to the challenge of over-segmentation.

\begin{figure}[!htbp]
    \centering
    \includegraphics[width=0.99\linewidth]{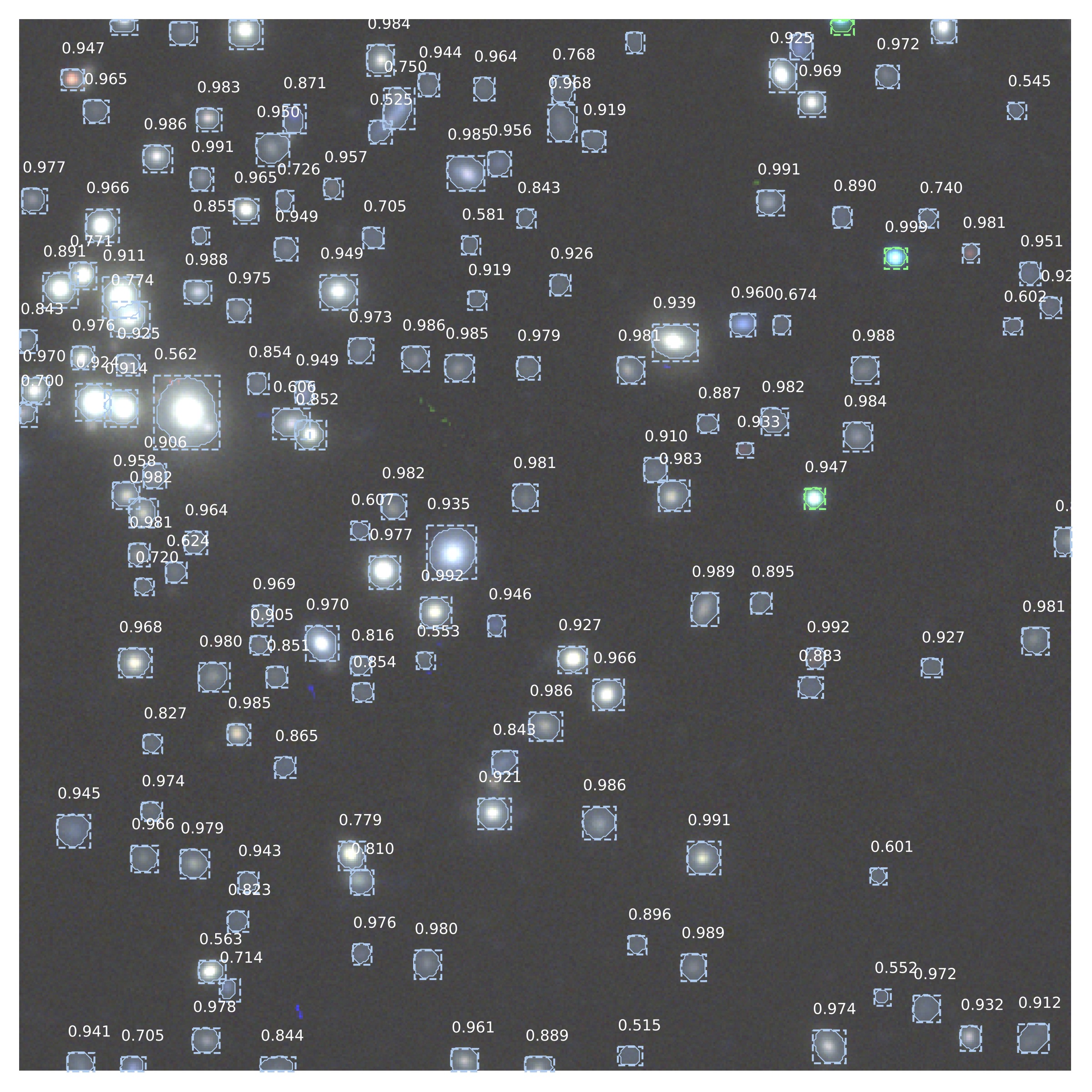}
    \caption{From \citet{2019MNRAS.490.3952B}, illustration of detection inference in an actual DECaLS image of ACO 1689. Galaxy masks are depicted in light blue, while star masks are presented in green. The confidence of the detection, indicating the likelihood that the object belongs to a specific class, is displayed above each mask.}
    \label{fig_instance_core}
\end{figure}

In the domain of astronomy, Mask R-CNN has found widespread application. It has been utilized for tasks such as segmenting solar filaments in H-$\alpha$ images of the Sun \citep{ahmadzadeh2019toward}, distinguishing different types of galaxies from SDSS data \citep{2020A&C....3300420F}, and efficiently detecting and segmenting stars and galaxies \citep{2019MNRAS.490.3952B}. Figure~\ref{fig_instance_core} provides an illustration of how Mask R-CNN identifies and segments stars and galaxies in real observational data.

As a side note, there were earlier Fully Convolutional Networks (FCN) based methods that predated Mask R-CNN, although Mask R-CNN has largely replaced these simpler FCN-based methods in image segmentation. These earlier FCN-based approaches employed straightforward Convolutional Neural Networks (CNNs) like AlexNet \citep{NIPS2012c399862d}, VGGNet \citep{simonyan2014very}, GoogLeNet \citep{szegedy2015going}, and ResNet \citep{he2016deep} as backbones, predicting class labels for each pixel in an image. For instance, \citet{2022MNRAS.511.5032B} utilized popular image classification model architectures such as VGG16 \citep{simonyan2014very}, ResNet50v2 \citep{he2016deep}, and Xception \citep{chollet2017xception} as backbones, incorporating saliency maps to highlight significant regions for segmentation of various components of galaxies. Building upon this eXplainable Artificial Intelligence (XAI) technique, \citet{tang2023model} further developed and enhanced it for the classification and segmentation of different types of \xd{radio} galaxies. Additionally, \citet{richards2023panoptic} combined both semantic segmentation and instance segmentation, fusing Mask R-CNN and FCN, to create a panoptic segmentation model for segmenting galactic structures.

\subsection{Self-Organizing Map}

\begin{figure}[!htbp]
    \centering
    \includegraphics[width=0.99\linewidth]{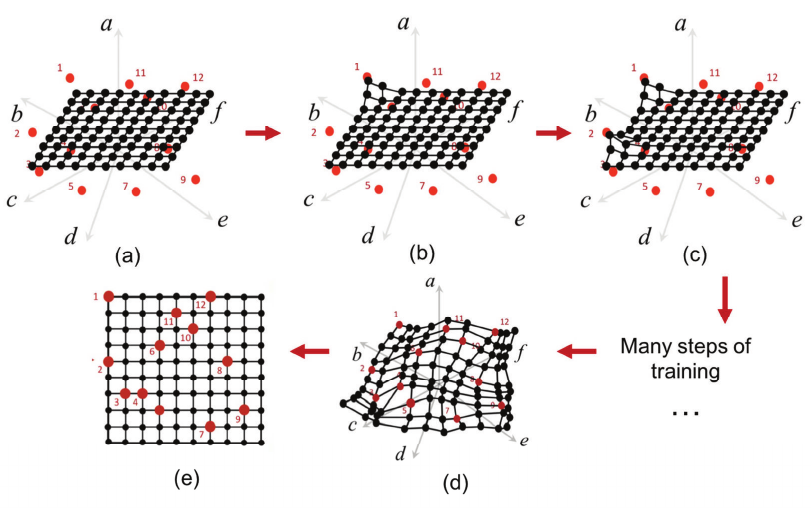}
    \caption{From \citet{qian2019introducing}, an illustration depicting the training of SOM with red dots representing training data in high-dimensional space, and black dots alongside the grid indicating the trained SOM.}
    \label{fig_som_19}
\end{figure}

Self-Organizing Map (SOM), also known as Kohonen maps, is a neural network-based technique rooted in the principles of unsupervised autonomous learning inspired by neurobiological studies \citep{kohonen1990self}. The SOM algorithm adapts to data through synaptic plasticity, mirroring the organized mapping of sensory inputs in the cerebral cortex. This topographic map maintains two critical properties: firstly, it retains incoming information in its proper context at each processing stage, and secondly, it ensures that neurons handling closely related information are positioned near one another, fostering interaction through short synaptic connections.

\begin{figure*}[!htbp]
    \centering
    \includegraphics[width=0.99\linewidth]{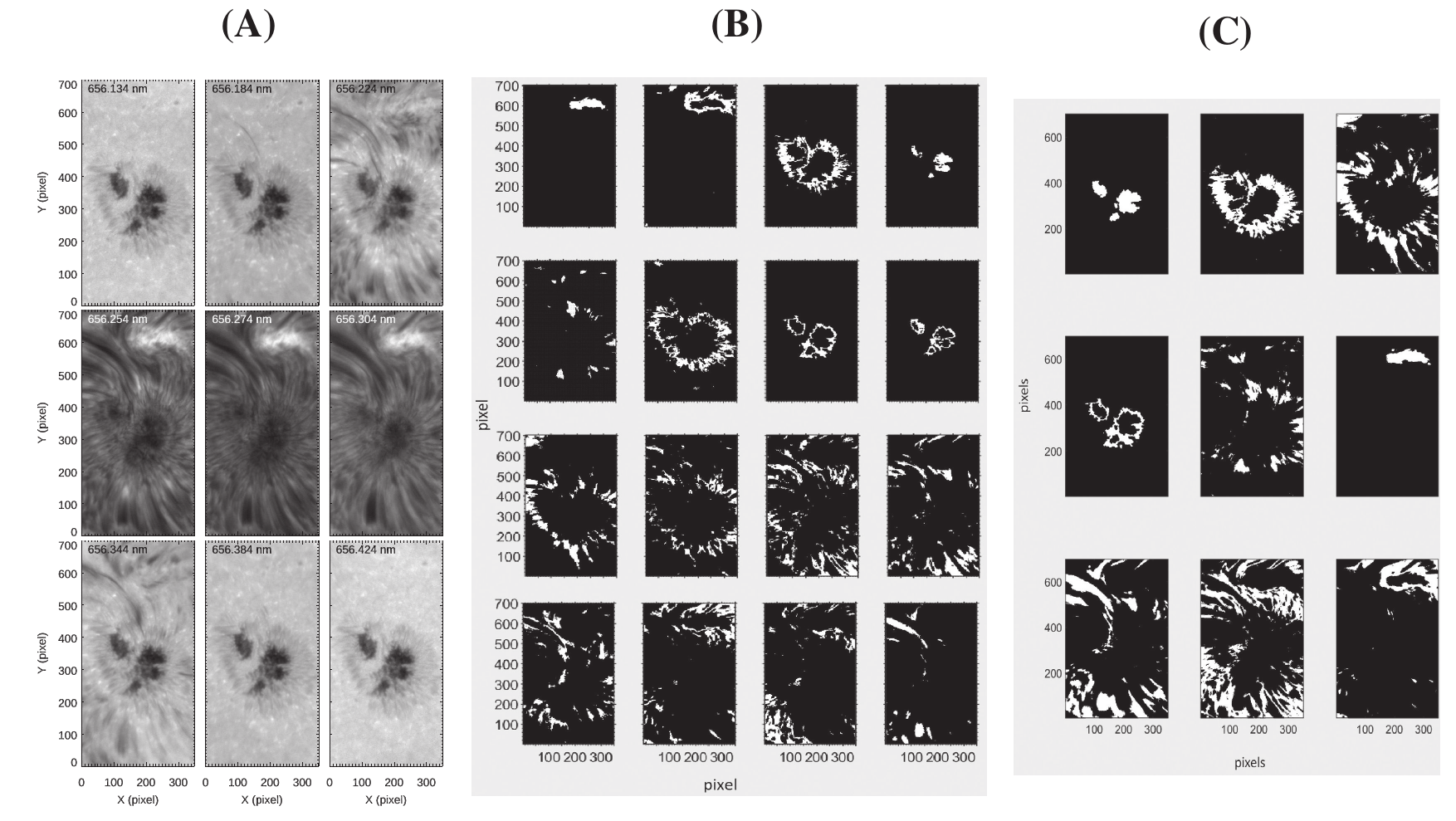}
    \caption{From \citet{2021MNRAS.503.2676S}, an illustration of the segmented output using SOM for a solar image. (A): A selection of spectral images obtained along the H$\alpha$ line. (B): The feature lattice generated by a $4\times 4$ SOM, facilitating the segmentation of 16 distinct regions. (C): The feature lattice produced by a $3\times 3$ SOM, enabling the segmentation of nine different regions.}
    \label{fig_som_solar}
\end{figure*}

In the realm of image segmentation, SOM stands out as a potent unsupervised learning algorithm, adept at discerning intricate patterns and structures within complex datasets. Its ability to map high-dimensional input data onto a lower-dimensional grid ensures the preservation of topology and relationships among data points, making it particularly well-suited for tasks such as image segmentation, where understanding spatial relationships is crucial. SOM excels in grouping similar pixels, facilitating the identification of distinct regions or objects. Its departure from conventional methods, which often rely on predefined criteria, is notable, as SOM learns patterns directly from the input data, offering an adaptive and data-driven approach to segmentation. The algorithm's versatility in capturing both global and local features positions it as an effective tool across a spectrum of image segmentation tasks.

Figure~\ref{fig_som_19} visually illustrates the training process of SOM. In addition to the introductory steps provided earlier in this section, we delve into the fundamental components and central procedures of SOM to provide a comprehensive understanding:
\begin{enumerate}

\item SOM Initialization: Configure the SOM grid by determining its size and dimensions. Initialize the SOM nodes with random weights.

\item Competition: Present each input image to the SOM and identify the winning node, which possesses weights most similar to the input. Update the weights of the winning node and its neighboring nodes using a learning rate and a neighborhood function. Repeat this process for multiple iterations, allowing the SOM to adapt to the input data.

\item Quantization: Assign each input image to the node with the closest weights, effectively mapping the high-dimensional data onto the lower-dimensional SOM grid.

\item Clustering and Labeling: Group similar nodes together to form clusters, representing distinct regions or structures in astronomical images. Analyze the clustered map and assign labels to different regions or clusters based on the characteristics of the astronomical objects they represent.
\end{enumerate}

SOM's inherent strength lies in its adept handling of complex, high-dimensional data, such as multi-band or hyperspectral images prevalent in remote sensing and astronomical observations. Its robustness against noise and variations in image intensity further enhances its efficacy in real-world applications. The SOM technique offers a unique and adaptive approach to image segmentation, proving to be a valuable tool for extracting meaningful structures and patterns from various types of imagery. This versatility and effectiveness extend across different domains, including astronomy, medical imaging, and remote sensing, showcasing its applicability to a wide range of segmentation challenges. As an illustration, \citet{2021MNRAS.503.2676S} employ SOM in astronomy to analyze a high spatial and spectral resolution H$\alpha$ line image of the sun, identifying several features corresponding to the main structures of the solar photosphere and chromosphere. Figure~\ref{fig_som_solar} provides an example of the segmented output generated by SOM for solar images.

\subsection{Encoder-Decoder Architectures}

Encoder-decoder architectures, a powerful paradigm in image segmentation, have revolutionized astronomy by enabling precise delineation of celestial objects and regions within astronomical images, a crucial step in understanding and analyzing astronomical structures. Encoder-decoder architectures comprise two key components:
\begin{enumerate}
\item[\large\textbullet] Encoder: The encoder extracts informative features from the input image using a sequence of convolutional layers that reduce spatial dimensions while capturing high-level features. In astronomy, the encoder learns to discern and learn from the intricate structures, objects, and phenomena present in astronomical images, even in noisy or complex backgrounds.

\item[\large\textbullet] Decoder: The decoder takes the encoded features and generates a segmentation mask using transposed convolutions to upsample the spatial dimensions and produce a high-resolution mask. In astronomy, the decoder translates the learned features into precise pixel-level delineation of astronomical objects, providing insights into their size, shape, and spatial distribution.

\end{enumerate}

Encoder-decoder architectures have evolved to include a diverse range of models, each with unique features and adaptations. These architectures, including U-Net~\citep{ronneberger2015u}, U-Net++~\citep{zhou2018unet++}, TransUNet~\citep{chen2021transunet}, and Swin-Unet~\citep{cao2022swin}, share a common overarching structure while diverging in their specific encoder and/or decoder implementations. The fundamental architecture provides a robust foundation, while the variations in implementation cater to specific segmentation needs and challenges. In subsequent sections, we explore the specific details of these encoder-decoder architectures, unraveling their unique attributes and contributions to image segmentation in astronomy and related domains.

\subsubsection{U-Net}

\begin{figure}[!htbp]
    \centering
    \includegraphics[width=0.48\textwidth]{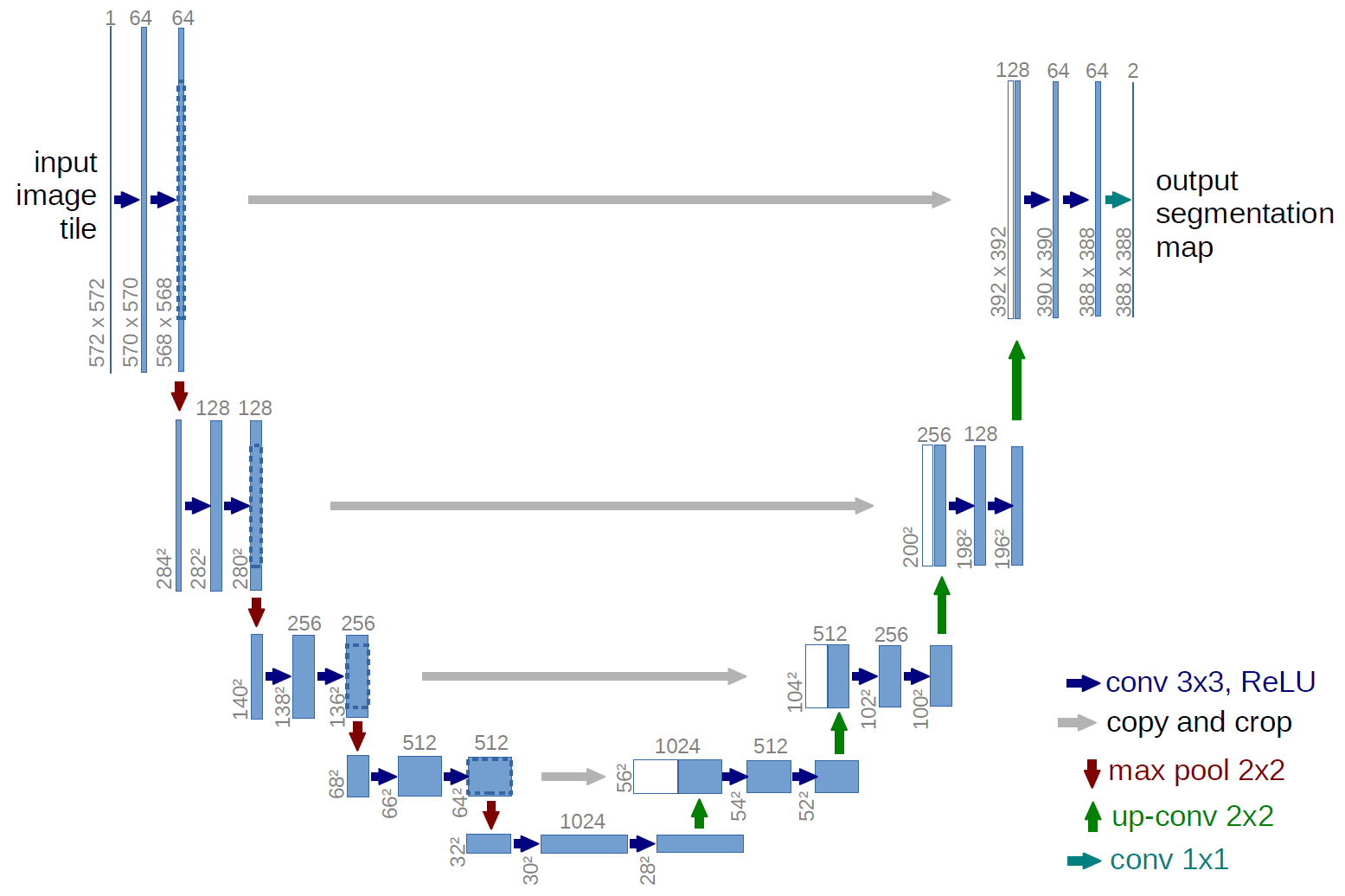}
    \caption{From \citet{ronneberger2015u}, a schematic representation of the U-Net architecture.}
\label{fig:unet}
\end{figure}

\begin{figure}[!htbp]
    \centering
    \includegraphics[width=0.99\linewidth]{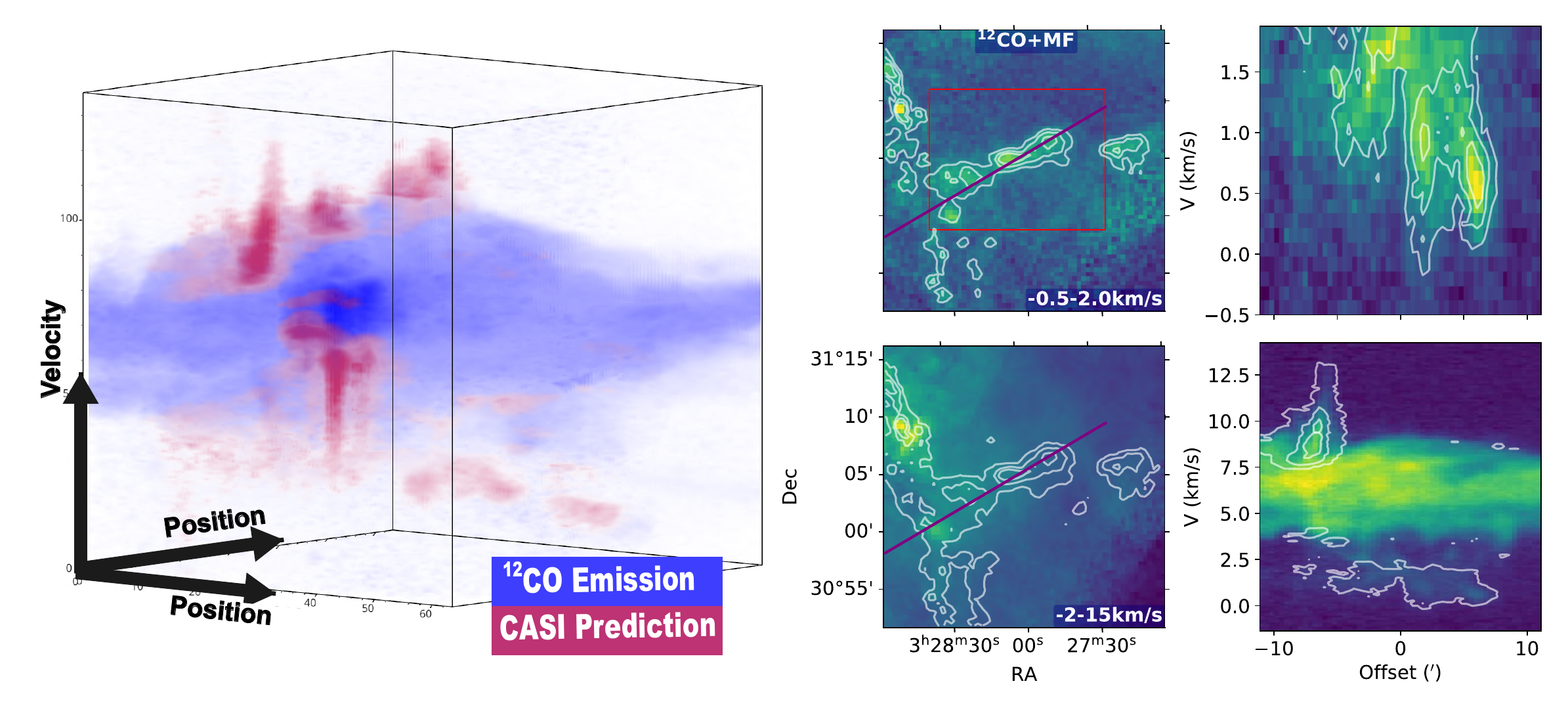}
    \caption{From \citet{2020ApJ...905..172X}. 3D visualization of the \CASItD\ prediction on the location of outflows.}
    \label{fig_proposal_plot_casi3d}
\end{figure}

U-Net~\citep{ronneberger2015u}, depicted in Figure~\ref{fig:unet}, is a pivotal encoder-decoder architecture that has significantly transformed image segmentation, including in astronomy. Its symmetrical structure with skip connections enables precise object delineation, even in complex and noisy backgrounds. This adaptability has found \xd{some} applications in astronomy, playing a pivotal role in tasks such as segmenting large-scale structures in simulations \citep{2019MNRAS.484.5771A}, outlining stellar wind-driven bubble structures in simulations \citep{2019ApJ...880...83V}, pinpointing stellar feedback structures in observational data \citep{2020ApJ...890...64X,2020ApJ...905..172X}, precisely capturing the intricate details of spiral arms in disk galaxies \citep{2021A&A...647A.120B}, and delineating galactic spiral arms and bars \citep{2023arXiv231202908W}. Moreover, it has proven effective in segmenting individual galaxies within cosmological surveys \citep{2020MNRAS.491.2481B,2021arXiv211115455B}, as well as in segmenting galaxy-galaxy strong lensing systems \citep{2022A&A...657L..14O} and locating subhalos from strong lens images \citep{2022ApJ...927...83O}. Figure~\ref{fig_proposal_plot_casi3d} visually depicts the 3D output of the Convolutional Approach to Structure Identification - 3D (\CASItD) prediction \citep{2020ApJ...905..172X}. Employing a U-net architecture, this method identifies the positions of protostellar outflows within a real 3D position-position-velocity data cube of the Perseus molecular cloud.

U-Net offers several advantages for image segmentation. Its versatility enables it to handle structures of diverse sizes and complexities, while its symmetrical architecture and effective use of skip connections preserve fine spatial details and facilitate efficient segmentation of objects of varying scales. This adaptability is crucial for working with astronomical images, which often feature astronomical objects spanning a wide spectrum of sizes and shapes.

However, U-Net also presents some limitations, primarily associated with its computational demands. Training and inference with U-Net models can be computationally intensive, requiring access to robust hardware resources. Achieving optimal performance with U-Net typically involves meticulous hyperparameter tuning, a process that can be time-consuming and iterative. Additionally, effectively deploying U-Net may necessitate a deep understanding of deep learning concepts. Finally, U-Net can be prone to overfitting on small datasets without the proper application of regularization techniques, potentially leading to suboptimal generalization when applied to new, unseen data.

\subsubsection{U-Net++}

\begin{figure}[!htbp]
    \centering
    \includegraphics[width=0.48\textwidth]{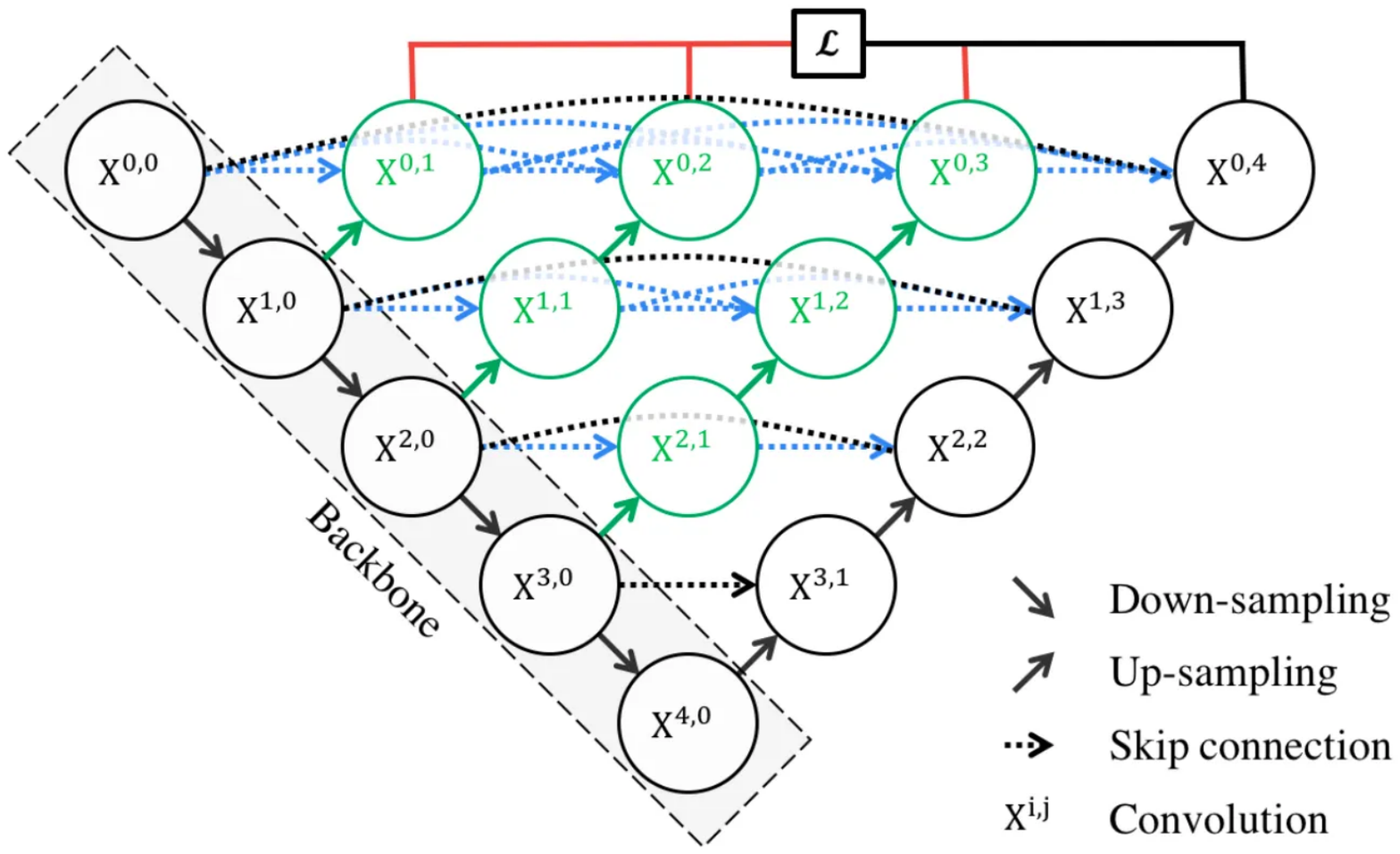}
    \caption{Overview of the U-Net++ architecture, adapted from \citet{zhou2018unet++}.}
\label{fig:unet++}
\end{figure}

\begin{figure}[!htbp]
    \centering
    \includegraphics[width=0.99\linewidth]{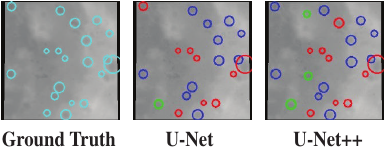}
    \caption{From \citet{Jia9380415}, moon crater detection results of different networks. Newly predicted craters are indicated by green circles, accurately recognized craters are represented by blue circles, and red circles signify unrecognized craters predicted by the network.}
    \label{fig_moon_jia2021}
\end{figure}

U-Net++~\citep{zhou2018unet++}, a refined version of the U-Net architecture illustrated in Figure~\ref{fig:unet++}, is specifically engineered to boost feature extraction and enhance segmentation precision. This is accomplished by incorporating nested dense convolutional blocks, which have the ability to seize a more extensive range of contextual information. This characteristic is especially useful when dealing with the intricacies of astronomical images, which frequently present a multitude of complex structures that necessitate accurate delineation. U-Net++ has been successfully utilized to segment filamentary structures within the interstellar medium with high precision in \citet{2023A&A...669A.120Z}, as well as in detecting and segmenting moon impact craters on the lunar surface \citep{Jia9380415}. Figure~\ref{fig_moon_jia2021} depicts exemplar outcomes of moon crater detection achieved using U-net and U-Net++ architectures.

When juxtaposed with its predecessor, U-Net, the advantages of U-Net++ become more pronounced. The inclusion of nested dense convolutional blocks within U-Net++ marks a substantial advancement. These blocks equip the model with the ability to access and integrate a more extensive set of contextual information, thereby fostering a deeper understanding of complex astronomical structures. This feature enables U-Net++ to attain more precise and accurate segmentation, which is of paramount importance in the realm of astronomy, where objects and structures can be highly complex and exhibit considerable variation in size and shape. U-Net++ has demonstrated its dominance by consistently surpassing U-Net in various medical image segmentation challenges, thereby cementing its status as the optimal choice for these tasks.

\subsubsection{TransUNet}

\begin{figure}[!htbp]
    \centering
    \includegraphics[width=0.48\textwidth]{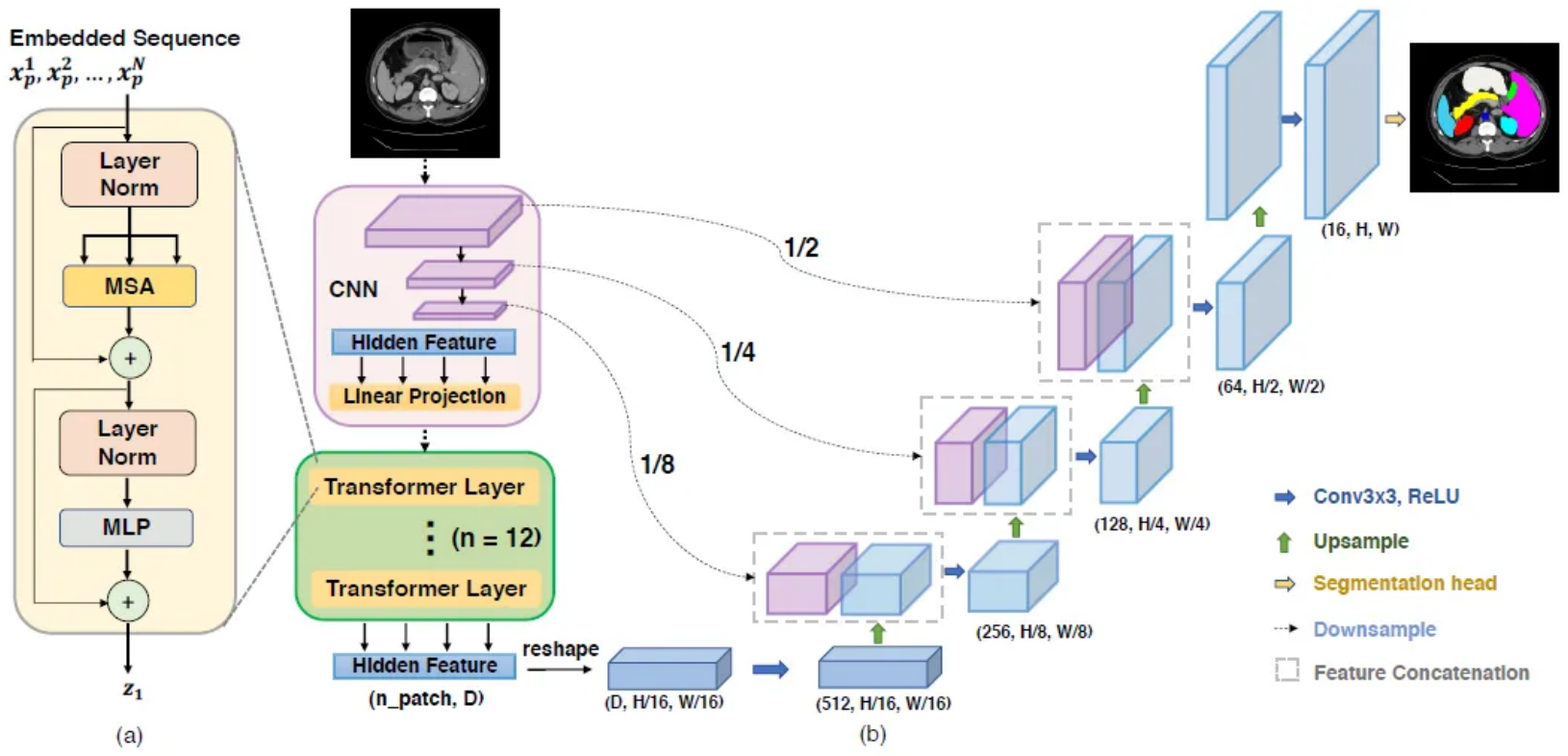}
    \caption{Overview of the TransUNet framework, as presented in \citet{chen2021transunet}.}
\label{fig:TransUNet}
\end{figure}

\begin{figure}[!htbp]
    \centering
    \includegraphics[width=0.99\linewidth]{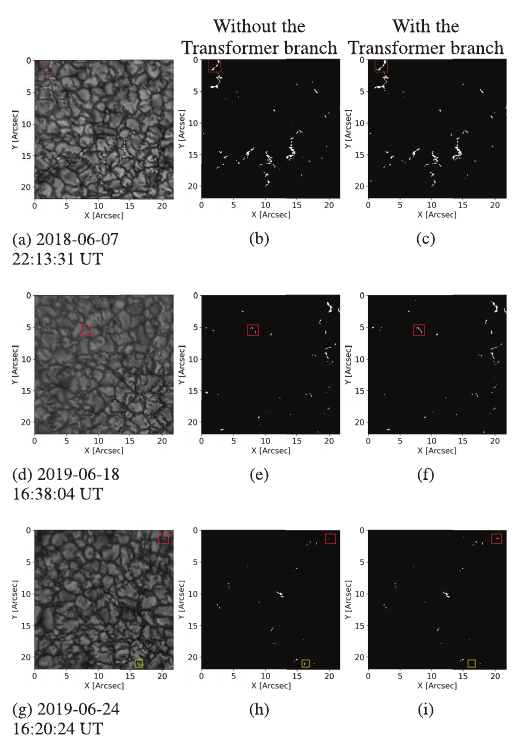}
    \caption{From \citet{2023A&A...677A.121Y}, segmentation results with and without the Transformer branch. The first column corresponds to the input image, the second column showcases the segmentation result without the Transformer branch, and the third column exhibits the segmentation result obtained using the Transformer branch. The red boxes highlight examples where the utilization of a dual-branch network enhances feature extraction precision and improves MBP segmentation. The yellow boxes point out instances where the dual-branch network leads to fewer misidentifications of MBPs.}
    \label{fig_transunet_solar}
\end{figure}

TransUNet~\citep{chen2021transunet}, depicted in Figure~\ref{fig:TransUNet}, is a novel combination of the transformer architecture and the encoder-decoder structure inherent to U-Net. This integration harnesses the powerful attention mechanism intrinsic to transformers, equipping the model with the ability to capture extensive dependencies across the entire dataset. This feature is especially beneficial when dealing with complex structures or objects that may cover large areas within an image. By combining the advantages of both transformers and U-Net, TransUNet presents a robust solution for tackling image segmentation challenges in astronomy, ensuring accurate delineation of celestial objects. \citet{Jia...AETransUNet} utilize TransUnet to extract small impact crater features on the moon with superior accuracy compared to other models. Additionally, \citet{2023A&A...677A.121Y} apply TransUnet to segment magnetic bright points (MBP) in the solar photosphere, achieving elevated accuracy. Figure~\ref{fig_transunet_solar} illustrates an instance of applying TransUnet to segment MBP in the solar photosphere, comparing the results with and without the Transformer branch.

Choosing between U-Net, U-Net++, and TransUNet hinges on the unique characteristics of the segmentation task and the inherent nature of the input data. While TransUNet presents distinct advantages in certain aspects, the comparison of performance is contingent on the complexity of the dataset and the specific demands of the segmentation task at hand.



\subsubsection{Swin-UNet}
\begin{figure}[!htbp]
    \centering
    \includegraphics[width=0.48\textwidth]{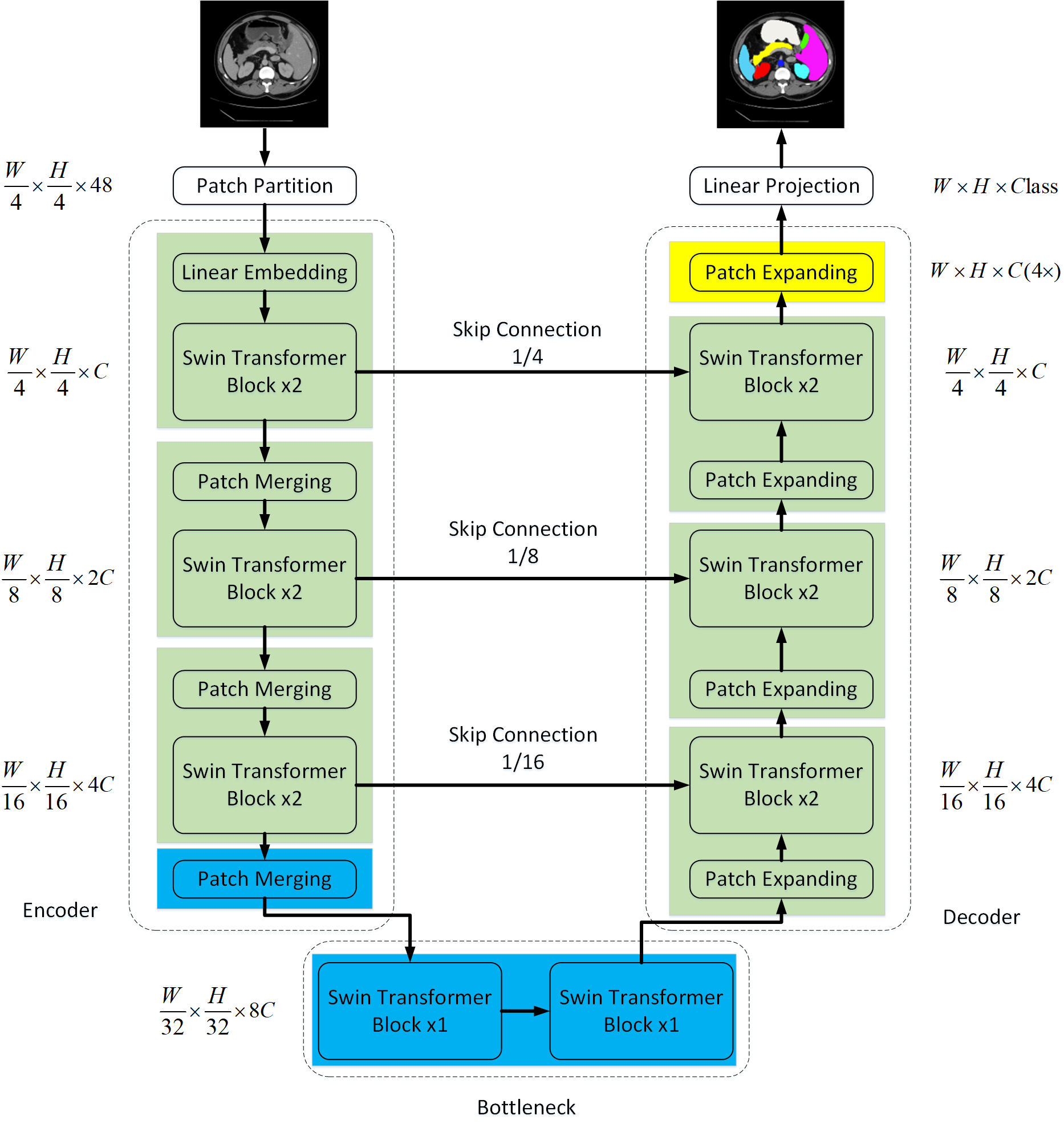}
    \includegraphics[width=0.48\textwidth]{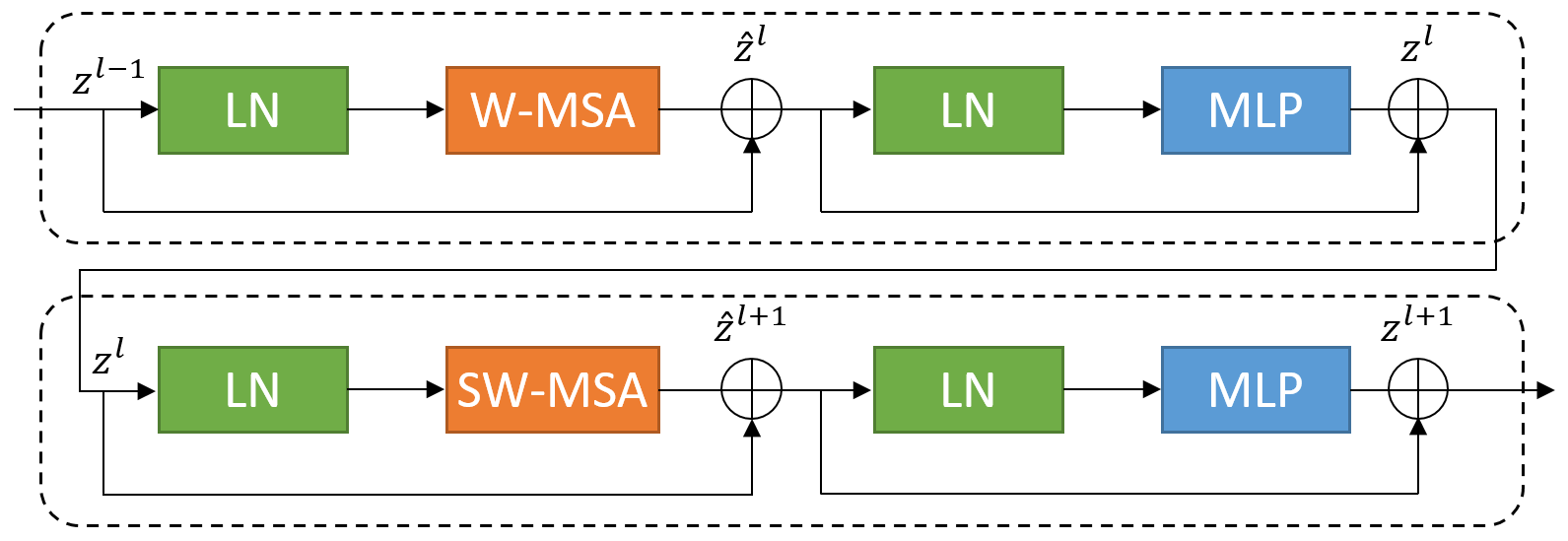}
    \caption{Overview of the \xd{Swin-UNet} framework, adapted from \citet{cao2022swin}.}
\label{fig:SwinUNet}
\end{figure}

Swin-UNet~\citep{cao2022swin}, depicted in Figure~\ref{fig:SwinUNet}, represents an innovative amalgamation that integrates the hierarchical architecture of the Swin Transformer with the encoder-decoder framework of U-Net. Swin Transformers are renowned for their ability to process images with diverse structures and scales efficiently. Swin-UNet leverages this efficiency to skillfully capture complex structures that may vary significantly in size. In situations where a single-scale approach may fail to accurately segment objects within astronomical images, Swin-UNet provides an effective solution.

\begin{figure}[!htbp]
    \centering
    \includegraphics[width=0.99\linewidth]{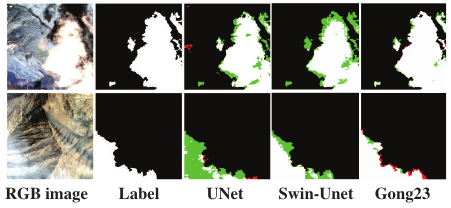}
    \caption{From \citet{rs15215264}, prediction outcomes of various models on the AIR-CD remote sensing dataset. The Gong23 model integrates both Swin-UNet and traditional CNN in its architecture.}
    \label{fig_swinunet_gong23}
\end{figure}

Swin-UNet has proven its efficacy in effectively handling complex and extensive medical images, substantially improving the precision of image segmentation tasks within the medical domain. While the utilization of Swin-UNet in astronomical contexts is somewhat limited, there are instances where it has been applied. For example, it has been employed in segmenting clouds from remote sensing images \citep{rs15215264} and detecting astronomical targets from multi-color photometry sky surveys \citep{JIA2023100687}. Figure~\ref{fig_swinunet_gong23} illustrates Swin-UNet's proficiency in accurately segmenting clouds from a remote sensing image.



\subsection{Vision Transformers (ViT)}

\begin{figure}[!htbp]
    \centering
    \includegraphics[width=0.48\textwidth]{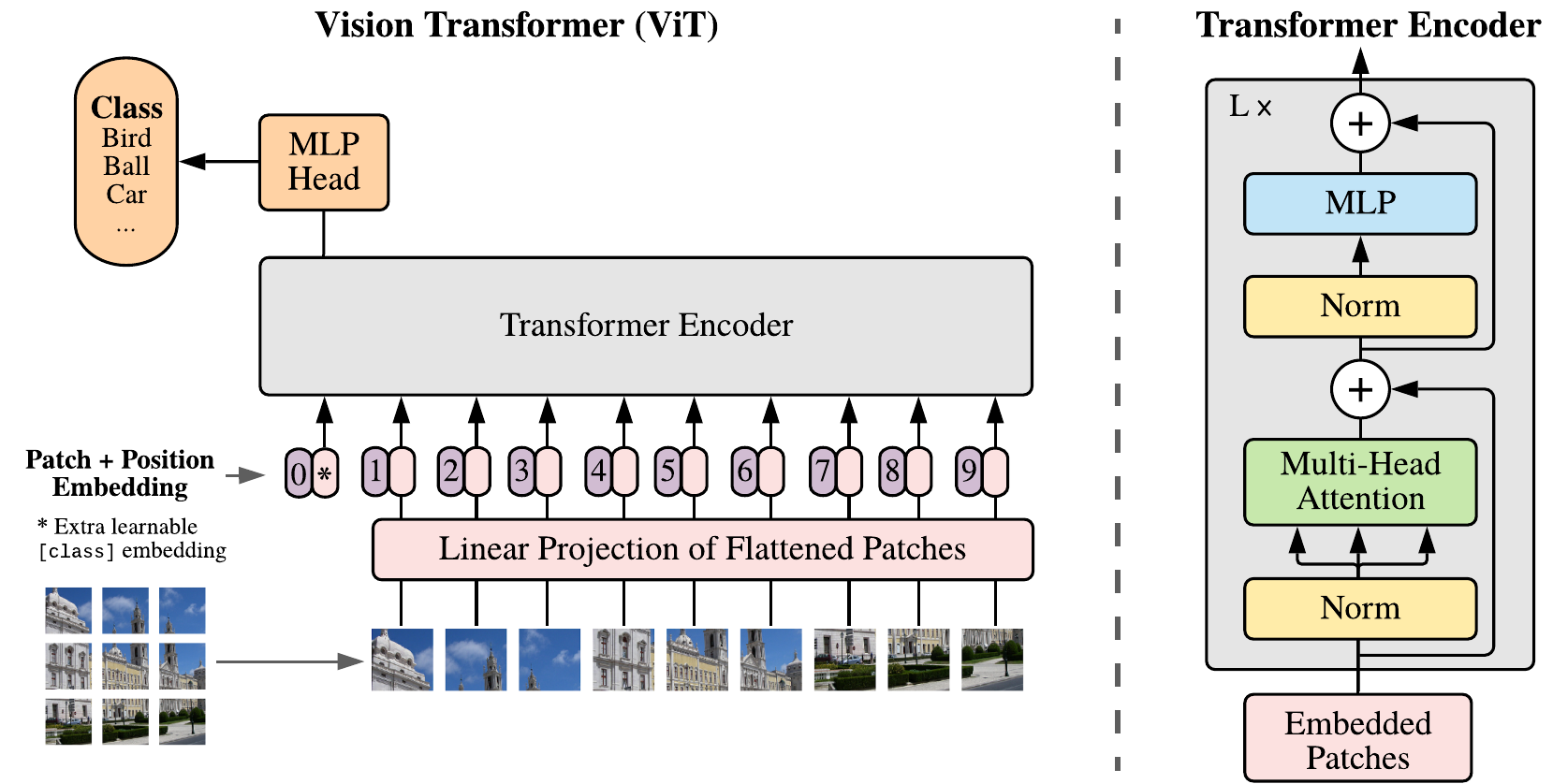}
    \caption{Overview of the Vision Transformer (ViT) framework as presented in \citep{dosovitskiy2020image}.}
\label{fig:ViT}
\end{figure}

The Vision Transformer (ViT) framework, illustrated in Figure~\ref{fig:ViT} and introduced by \citet{dosovitskiy2020image}, represents an innovative deep learning architecture that has transformed the landscape of image segmentation. ViTs leverage the power of transformers \citep{vaswani2017attention}, originally designed for natural language processing (NLP) tasks, to capture long-range dependencies and intricate spatial relationships within images. At the heart of transformers is the concept of attention, which allows them to focus on specific regions of an image, leading to highly effective feature extraction. The integration of ViTs with image segmentation has ushered in a groundbreaking paradigm for comprehending and delineating objects within scientific imagery, notably in the context of medical images. The potential for applying this innovative approach to astronomy images in the future holds great promise, offering the prospect of achieving even greater precision and adaptability in segmentation results.

To understand the capabilities of ViTs in image segmentation, it is essential to explore the foundations of transformers and attention mechanisms. Transformers are a class of deep learning models that have revolutionized various domains, including NLP and computer vision. These models rely on a mechanism called ``attention," which enables them to process sequential or spatial data effectively. In the context of computer vision, transformers break down images into smaller patches, treating them as sequences of data. This approach allows them to capture global context and intricate spatial relationships, which are vital for tasks like image segmentation.

Attention, the key mechanism within transformers, enables the model to assign varying levels of importance to different parts of the input data. By learning these importance weights, the model can focus on relevant information while filtering out noise or irrelevant details. This mechanism's ability to capture long-range dependencies in the data has made transformers, and subsequently ViTs, exceptionally effective in image segmentation tasks. In astronomy, where celestial objects can vary in size, shape, and distribution, the power of transformers and their attention mechanisms can significantly enhance the accuracy and robustness of segmentation processes.

In addition to the previously outlined general steps in this section, we explore the fundamental components and central procedures of ViTs to offer a comprehensive understanding:
\begin{enumerate}

\item Image Patching: ViTs divide input images into smaller, non-overlapping patches. This process transforms the 2D image data into a sequence of 2D patches. These patches serve as the input to the ViT model. Patch size is a crucial hyperparameter to consider, as it affects the trade-off between computational efficiency and capturing fine-grained details.

\item Embedding the Patches: Each patch is flattened into a 1D vector and linearly projected to create embeddings. These embeddings carry information about the patches and are the input data for the ViT model. They allow ViTs to work with sequences of data, which is a fundamental concept in NLP and now applied to images.

\item Positional Encoding: Since ViTs lack the inherent spatial understanding of CNNs, they incorporate positional encoding to provide spatial information to the model. This encoding informs the model about the relative locations of patches within the image. Various positional encoding techniques, including sinusoidal encoding or learned positional encodings, can be used.

\item Design Model Architecture: Design a ViT architecture for image segmentation. ViTs, unlike CNNs, use a transformer architecture. A typical ViT model comprises several key components:\begin{itemize}
    \item Embedding Layer: This layer converts image patches into embedding vectors.
    \item Transformer Encoder Blocks: These blocks process the embeddings and capture spatial relationships and context information across patches.
    \item Class Token: ViTs add a class token to the embeddings to perform classification.
    \item Positional Encodings: These encodings help the model understand the spatial position of patches.
    \item Linear Projection: This projection maps the transformer output to the segmentation mask space.   
    \end{itemize}

\end{enumerate}

ViTs have become increasingly prominent in the field of computer vision, demonstrating remarkable proficiency, particularly in scientific image segmentation. In domains like medical image segmentation, including X-ray, CT, and MRI datasets, ViTs have showcased SOTA performance and remarkable accuracy, as summarized by \citep{henry2022vision}.

\begin{figure}[!htbp]
    \centering
    \includegraphics[width=0.99\linewidth]{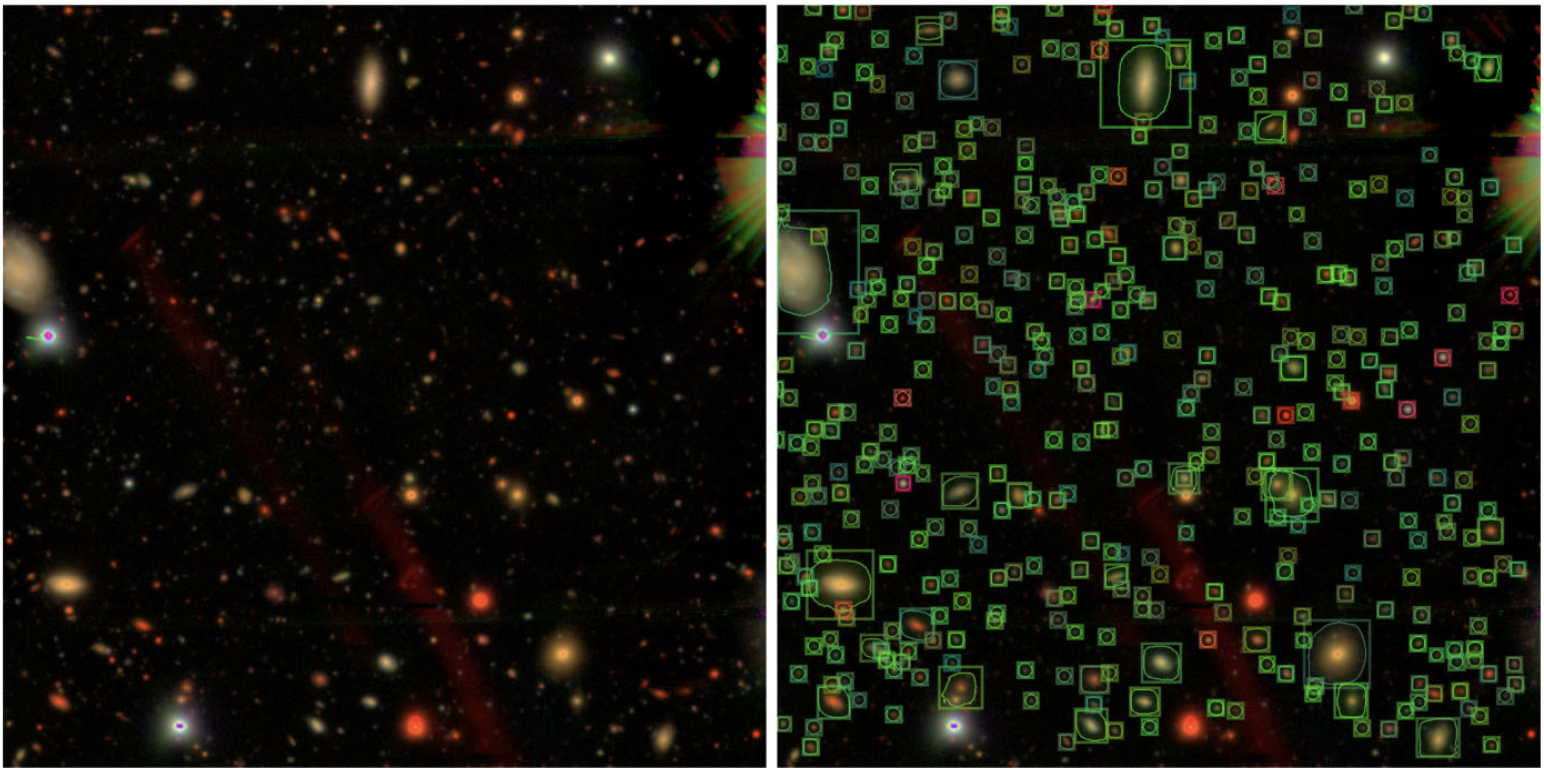}
    \caption{\xd{From \citet{2023MNRAS.526.1122M}: Left: Image showing artifacts like blooming and optical ghosts around the bright star in the upper right and large ghosts in the lower middle. Right: Inference results from an MViTv2 Lupton-scaled network.}}
    \label{Merz_MViT}
\end{figure}

\begin{figure}[!htbp]
    \centering
    \includegraphics[width=0.99\linewidth]{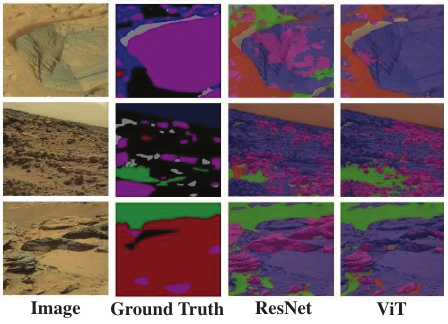}
    \caption{From \citet{rs14246297}, a comparative analysis of terrain segmentation on Mars using ResNet and ViTs.}
    \label{fig_vit_mars}
\end{figure}

\xd{There are pioneering direct applications of ViTs in astronomy for segmentation tasks, as demonstrated by \citet{2023MNRAS.526.1122M} using MViTv2 (Multiscale Vision Transformers) for segmenting astronomical objects, and by \citet{rs14246297} applying ViTs for terrain segmentation on Mars. Figure~\ref{Merz_MViT} presents an example of MViTv2 used to segment astronomical objects in an image containing artifacts like blooming and optical ghosts, with the model effectively ignoring these artifacts in its predictions. For a visual comparison across different models, Figure~\ref{fig_vit_mars} illustrates the performance of ViTs in segmenting terrain on Mars.} Beyond segmentation, ViTs have also ventured into various astronomical applications. For instance, they have been employed in tasks such as classifying transient astronomical sources \citep{2023arXiv230909937C} and estimating strong gravitational lensing parameters \citep{2022arXiv221004143H}. These initial explorations into the application of ViTs in astronomy represent an exciting frontier. ViTs bring the transformative capabilities of transformers and attention mechanisms to this domain, potentially reshaping the way we analyze and comprehend astronomical structures in the vast expanse of astronomical images in the future.

\subsection{Mamba}

\begin{figure}[!htbp]
    \centering
    \includegraphics[width=0.48\textwidth]{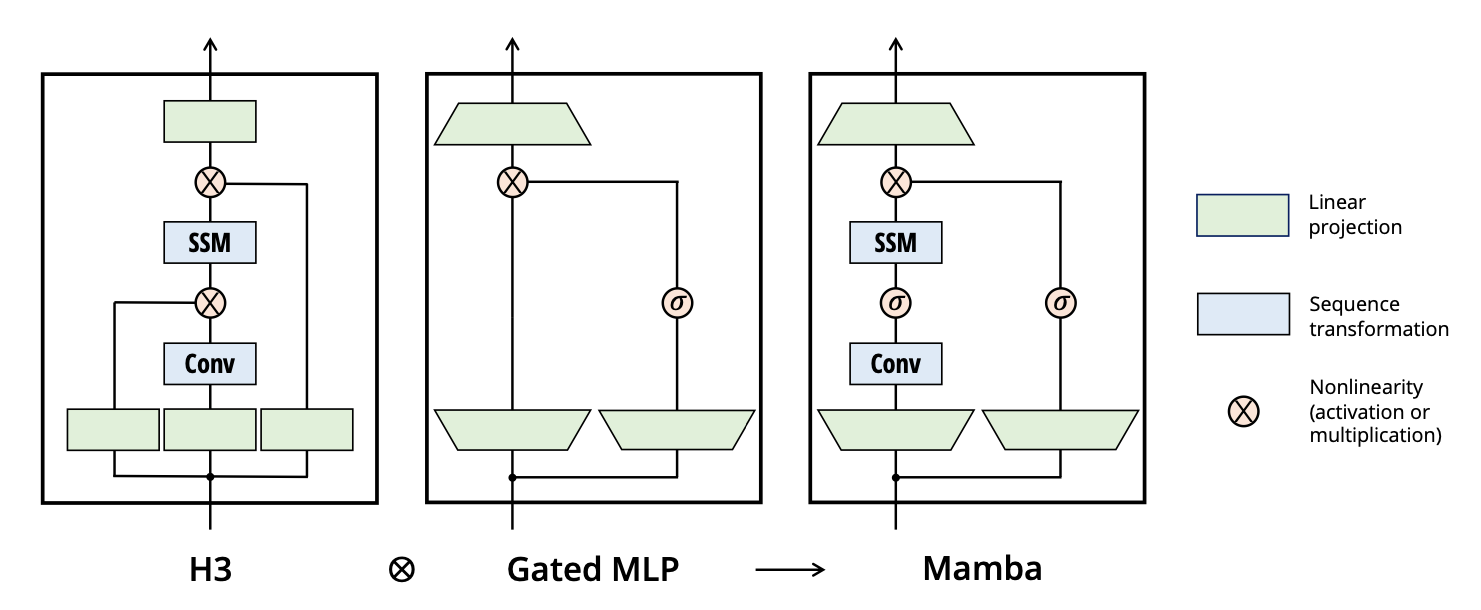}
    \caption{Overview of the Mamba framework as presented in \citep{gu2023mamba}.}
\label{fig:mamba}
\end{figure}

\begin{xdpar}
    
Mamba~\citep{gu2023mamba}, depicted in Figure~\ref{fig:mamba}, emerged as a recent model architecture in the machine learning community, showcasing strong performance across various downstream sequence modeling tasks. Originally designed for large-scale natural language processing tasks with extended sequence lengths akin to Transformers, Mamba introduces structured state space models (SSMs) and a hardware-aware parallel algorithm in recurrent mode. This design simplifies the traditional Transformer architecture, which heavily relies on attention mechanisms and MLP blocks, resulting in significantly faster inference speeds (5x higher throughput) and linear scalability with data sequence length.

One pivotal component of Mamba is its selective state space model (SSM) layer, which assigns different weights to inputs, enabling the model to prioritize predictive data for specific tasks. This adaptability allows Mamba to excel in various sequence modeling tasks, spanning languages, images, audio, and genomics.

While Mamba has not yet been applied to astronomical image segmentation, its proficiency in modeling variable and lengthy sequences suggests potential benefits for this community with straightforward adaptation. The data pre-processing pipeline for Mamba resembles that of Transformers but omits the requirement for positional encoding.

\end{xdpar}

\subsection{Generative Models}

\begin{figure}[!htbp]
    \centering
    \includegraphics[width=0.48\textwidth]{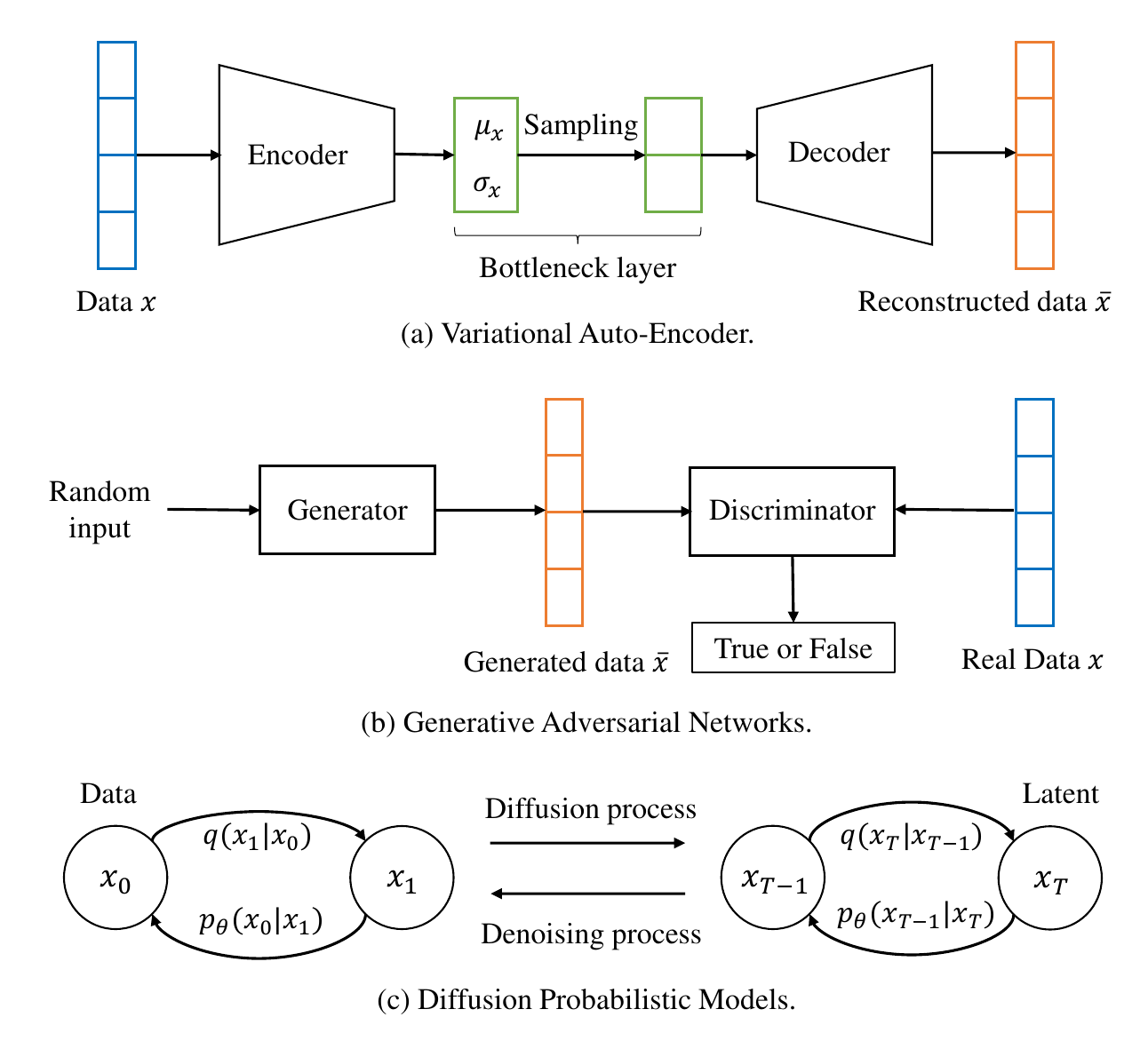}
    \caption{Overview of the general frameworks for three generative models, figure adapted from~\citep{zhu2022vision}.}
\label{fig:generative_models}
\end{figure}

Generative models constitute an intriguing subset of deep learning that transcends the confines of conventional image segmentation. They not only identify and outline objects within images, but also generate new data based on discerned patterns and structures. These models are engineered to grasp the inherent distribution of data, enabling them to generate novel, data-coherent content. In the realm of image segmentation, generative models offer a unique proficiency: they can fabricate detailed and lifelike segmentations of objects, a capability that can prove invaluable in various fields, including astronomy. Generative models, utilizing techniques like Variational Autoencoders (VAEs)\citep{vae-kingma2013auto}, Generative Adversarial Networks (GANs)~\citep{gan}, and Denoising Diffusion Probabilistic Models (DDPMs)~\citep{sohl2015dpm_thermo,ho2020dpm}, are capable of simultaneously identifying inherent patterns in an image and generating segmentations that mirror these patterns accurately. This vibrant fusion of image interpretation and generation paves the way for new possibilities in object identification and precise image outlining. The schematic representations of these generative models are depicted in Figure~\ref{fig:generative_models}. 

\begin{figure}[!htbp]
    \centering
    \includegraphics[width=0.99\linewidth]{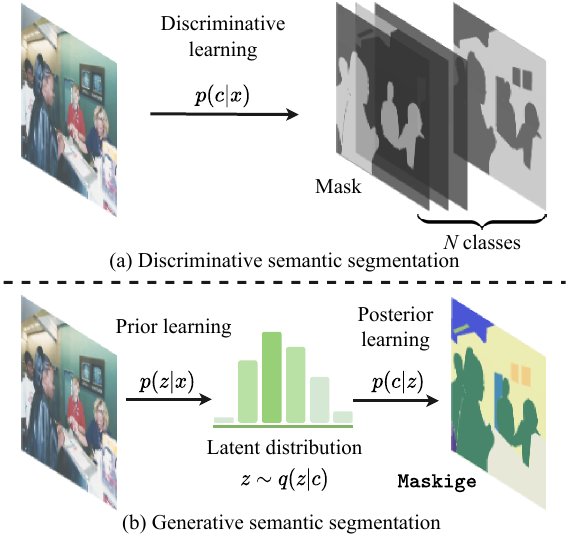}
    \caption{From \citet{chen2023generative}, a schematic comparison illustrating (a) conventional discriminative learning and (b) a generative learning-based model for segmentation.}
    \label{fig_intro_drawio}
\end{figure}

In contrast to conventional deep learning models for segmentation, which rely on the log-likelihood of a conditional probability (i.e., the classification probability of image pixels), generative segmentation models introduce an auxiliary latent variable distribution. This represents a notable departure from the conventional discriminative segmentation deep learning paradigm. The conditional generative model exhibits minimal requirements for task-specific architecture and loss function modifications, fully capitalizing on the capabilities of off-the-shelf generative models \citep{chen2023generative}. This characteristic positions them as a promising method in image segmentation. A schematic comparison between conventional discriminative learning and a generative learning-based model for segmentation is illustrated in Figure~\ref{fig_intro_drawio}.

\subsubsection{Variational Autoencoders (VAEs)}
Variational Autoencoders (VAEs)~\citep{vae-kingma2013auto} represent a powerful class of generative models used in various domains, inclding computer vision and image processing. These models excel at learning the underlying structure of data and, in the context of image segmentation, play a pivotal role in understanding image patterns and generating coherent segmentations. VAEs offer a unique approach to both data compression and generation, making them invaluable in tasks like image reconstruction and synthesis. 
VAE training usually seeks to minimize the Kullback–Leibler (KL) divergence (a statistical metric that measures the similarity between two distributions) via the Gaussian reparametrization in the bottleneck layer as part of the overall loss function. Additional image perception losses such as L1 are further introduced to ensure the reconstruction ability of the neural networks. The overall learning objective can be formulated as:
\begin{equation}
    \mathrm{log} \: p(x) = \mathbb{E}_{q(x|z)}[\mathrm{log}\:p(x|z)] - D_{KL}[q(z|x)||q(z)],
\end{equation}
where $p$ represents the decoder, $q$ is the encoder, $x$ and $z$ denote the original raw data and the learned latent embedding, respectively.

In addition to the general training steps outlined at the start of this section, we also provide a breakdown of key components and steps involved in training and operating VAE:
\begin{enumerate}

\item Model architecture design: A VAE consists of an encoder and a decoder. The encoder compresses the input image into a lower-dimensional latent space representation, while the decoder reconstructs the image from the latent representation.

\item Objective function: The VAE objective function consists of two parts: the reconstruction loss, which measures how well the VAE can reconstruct the input image, and the regularization term, which ensures that the latent space follows a desired distribution, such as a Gaussian distribution.

\item Training: The VAE is trained on the prepared dataset by minimizing the objective function. During training, the encoder learns to map input images to a structured latent space, while the decoder learns to generate accurate segmentations.

\end{enumerate}

VAEs offer several unique advantages for image segmentation. They are capable of both learning and generating segmentations effectively, due to their ability to create a structured latent space where images are represented in a continuous and meaningful way. This feature facilitates the capture of underlying patterns and structures within images. Additionally, VAEs can generate segmentations for previously unseen data, making them valuable for tasks involving new or unobserved data, such as in medical or astronomical imaging applications.

However, VAEs come with their set of limitations. Their complex architecture, with encoder, decoder, and regularization components, can be challenging to train and fine-tune, especially for beginners in deep learning. The regularization terms can also oversmooth segmentations, losing fine details. This can make VAEs unsuitable for tasks that require preserving intricate image elements, such as medical image segmentation. Training and running VAEs can be computationally expensive, especially for large or high-dimensional images. This can limit their practicality for users with limited computational resources. Additionally, although VAEs are relatively data-efficient, they still require a certain amount of labeled data for training. Acquiring a sufficiently large and diverse dataset for segmentation tasks can be challenging, especially in some domains.

\begin{figure}[!htbp]
    \centering
    \includegraphics[width=0.99\linewidth]{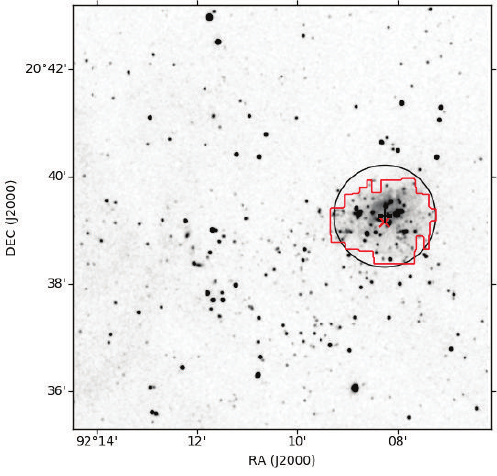}
    \caption{From \citet{Karmakar8634903VAE}, comparison between clusters detected using VAEs (highlighted in red) and detection results from \citet{2006A&A...452..203T} (depicted by black circles) around IRAS 06055+2039. The red cross indicates the detected center as per VAE, while the plus sign denotes the position of the IRAS point source.}
    \label{fig_vae_starcluster}
\end{figure}

Although VAEs have found \xd{some success in} applications in medical image segmentation, addressing challenges such as segmenting ambiguous medical images \citep{kohl2018probabilistic} and detecting incorrect segmentations in medical images \citep{sandfort2021use}, their application in segmentation within astronomy studies is relatively limited. Notably, \citet{Karmakar8634903VAE} applied VAEs and Gaussian Mixture Models in tandem to detect and segment stellar clusters from near-infrared data. Figure~\ref{fig_vae_starcluster} illustrates the performance of VAEs in segmenting stellar clusters in a near-infrared image.


\subsubsection{Generative Adversarial Networks (GANs)}

Generative Adversarial Networks (GANs)~\citep{gan} are a type of deep learning model that can generate new data, such as images or text, that is indistinguishable from real data. GANs are composed of two neural networks: a generator and a discriminator. The generator creates new data, while the discriminator attempts to distinguish between real and generated data.


GAN training operates on the premise that the discriminator should adeptly capture the characteristics of the target distribution. If a generator manages to deceive a well-trained discriminator, it signifies the generator's effectiveness as an approximator for the target distribution. The training process involves both the generator and discriminator working in tandem in an adversarial manner. Here, the generator strives to deceive the discriminator by producing more realistic data, while the discriminator endeavors to enhance its ability to distinguish between real and generated data. This adversarial competition propels both networks to refine their performance, leading to the generation of progressively more realistic data. In terms of loss functions, while the discriminator and generator are trained together, they are optimized using different loss functions in an adversarial way:
\begin{equation}
\begin{split}
    & \mathrm{min}_{G}\mathrm{max}_{D}\:V(D,G) = \\
    & \mathbb{E}_{x\sim p_{data}(x)}[\mathrm{log}D(x)] + \mathbb{E}_{z \sim p_Z(z)}[\mathrm{log}(1-D(G(z)))],
\end{split}
\end{equation}
where $G$ and $D$ are the generator and discriminator, respectively. $x$ and $z$ denotes the original raw data and the learned latent embedding, respectively.


GANs have revolutionized many fields, including computer vision, image processing, and natural language processing. They are used in a wide range of applications, including image generation, style transfer, and image segmentation.

In the context of image segmentation, GANs can be used to generate labeled images, which can be used to train machine learning models to perform segmentation tasks more accurately. GANs can also be used to create synthetic but realistic data for segmentation tasks in challenging domains, such as astronomy, where acquiring real labeled data can be difficult or expensive. 

Apart from the general training procedures introduced earlier in this section, we offer an in-depth exploration of the essential elements and processes required for training and utilizing GANs:
\begin{enumerate}

\item Designing the Network: GANs consist of two primary components---the generator and the discriminator. The generator employs random noise as input to generate images, while the discriminator's function is to differentiate between real and generated images.

\item Defining Loss Functions: The effectiveness of GAN training depends on carefully defining loss functions for the generator and discriminator. The generator seeks to minimize its loss function, which quantifies its ability to deceive the discriminator. Conversely, the discriminator aims to minimize its loss function, which measures its capability to distinguish between real and synthetic data.

\item Training Process: \xd{GAN training is an iterative adversarial process where the generator and discriminator compete to enhance their performances. Initially, the generator produces synthetic data evaluated by the discriminator. Subsequently, the generator updates its parameters to create more realistic data, while the discriminator adjusts its parameters to better discern between real and synthetic data. This iterative process continues until both networks converge to optimal states. Ideally, the generator's loss decreases while the discriminator's loss stabilizes. Notably, during discriminator training, the generator remains constant to enable the discriminator to distinguish real from synthetic data effectively. Conversely, during generator training, the discriminator remains constant to prevent the generator from chasing a moving target, thereby facilitating convergence and iterative improvement of both networks.}

\end{enumerate}

GANs can create high-quality synthetic data that is similar to real data. This is useful for image synthesis and segmentation. GANs can also augment existing datasets, reducing the need for manual data labeling. This can improve model performance in image segmentation. Additionally, GANs can generate realistic images, making them useful for image-to-image translation and image segmentation. GANs are versatile and can be adapted for various applications, including image segmentation in astronomy and medical imaging.

\begin{figure}[!htbp]
    \centering
    \includegraphics[width=0.99\linewidth]{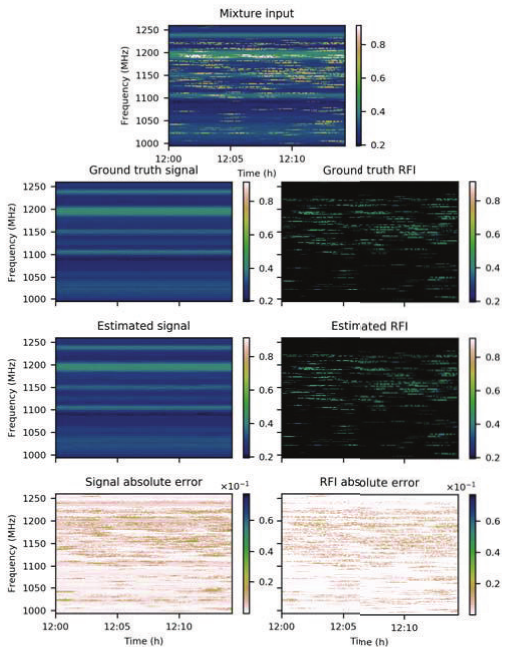}
    \caption{From \citet{Vos8918820GAN}, illustrations of the input mixture and the ground truth signal and RFI components (rows 1-2). Additionally, the separated components by GANs are presented, accompanied by the corresponding absolute error maps (rows 3-4).}
    \label{fig_GAN_rfi}
\end{figure}

However, GANs can be challenging to train. It can be difficult to achieve convergence and stability, and it often requires significant computational resources and expertise. GANs can also suffer from mode collapse, where they generate limited variations of data. GANs are sensitive to hyperparameter settings, and it can be time-consuming to select the right settings for learning rate, architecture, and batch size. Additionally, GANs can generate unrealistic artifacts or biases present in the training data. GANs also require a large amount of data for effective training. Finally, GANs are complex models, requiring expertise in deep learning for successful implementation and management.

Despite the inherent challenges, GANs showcase remarkable capabilities in data generation and refinement, particularly within the realm of image segmentation, presenting a potential revolution in the field. Notably, \citet{2021SoPh..296..176L} harnessed Conditional GANs to effectively segment filaments in low-quality solar images, surpassing the performance of traditional UNet methods. Additionally, \citet{2019MNRAS.485.2617R} utilized GANs to tackle the issue of deblending overlapping galaxies, a task comparable to galaxy segmentation in managing scenarios with overlapping galaxies. In the domain of radio data, \citet{Vos8918820GAN} utilized GANs for segmenting radio frequency interference (RFI). Figure~\ref{fig_GAN_rfi} illustrates an application of GANs in segmenting radio signals and RFIs from noisy radio data. These instances underscore the promising contributions of GANs to image segmentation, signaling their potential to reshape the landscape of this field significantly.

\subsubsection{Denoising Diffusion Probabilistic Models (DDPMs)}

Denoising Diffusion Probabilistic Models (DDPMs)~\citep{sohl2015dpm_thermo,ho2020dpm} are a class of advanced generative models that have emerged as a powerful approach in machine learning and image processing. They are designed to tackle challenging problems related to image generation, denoising, and restoration, including image segmentation.

DDPMs are inspired by the concept of diffusion, which is the process of gradually spreading noise throughout an image. In contrast, DDPMs work by iteratively removing noise from an image, gradually revealing the underlying structure. This iterative process enables DDPMs to effectively model the complex relationships between noisy and clean images, making them proficient in both denoising and generating realistic images. 
While there has been fast development regarding the diffusion training techniques tailored for various generative tasks in machine learning and computer vision~\citep{austin2021structured,rombach2022high,zhu2023discrete}, vanilla DDPMs are trained on a variational lower bound defined as follows:
\begin{equation}
     \begin{split}
     & \mathcal{L}_\mathrm{vb} = \mathbb{E}_q[\underbrace{D_\mathrm{KL}(q(x_T|x_0)||p(x_T))}_{\mathcal{L}_T} + \\
     & \sum_{t>1}\underbrace{D_\mathrm{KL}(q(x_{t-1}|x_t,x_0)||p_{\theta}(x_{t-1}|x_t))}_{\mathcal{L}_{t-1}}-\underbrace{\mathrm{log}\:p_{\theta}(x_0|x_1)}_{\mathcal{L}_0}],
     \end{split}
    \label{eq:vlb}
\end{equation}
where $q$ and $p$ represent the diffusion and denoising processes, respectively. $x_{i}$ denotes the data at diffusion step $t$. $\theta$ stands for learnable model parameters.

While they are quite advanced and require a deep understanding of probabilistic modeling and neural networks, here is a simplified overview of how they work:
\begin{enumerate}
\item  Forward Diffusion Process: The forward diffusion process starts with a clean image and gradually adds noise to it, until it reaches a desired level of noise. 

\item  Reverse Diffusion Process: The reverse diffusion process starts with a noisy image from the forward diffusion process and gradually removes noise from it, until it is as close to the original image as possible. This is done by reversing the steps of the forward diffusion process using a well-trained DDPM model.

\item  Image Generation: A trained DDPM model can be used to generate realistic images. To do this, a noisy or corrupted image is fed to the model. The model then denoises and rejuvenates the image to approximate the original image.

\item Image segmentation: In the context of image segmentation, conditional DDPM is often used. In this case, the original image is the conditional input, and the corresponding segmentation mask is the target. The goal is to recover the mask as accurately as possible through the reverse diffusion process. 

\end{enumerate}

DDPMs offer several advantages for image segmentation. First, they can generate high-quality and realistic images by effectively removing noise, which is essential for image segmentation. Second, DDPMs are designed to preserve the underlying structure and features of an image during denoising, which is important for accurate segmentation of object boundaries and intricate details. Third, DDPMs are based on a probabilistic framework, which allows them to model the complex relationships between noisy and clean images.

\begin{figure}[!htbp]
    \centering
    \includegraphics[width=0.99\linewidth]{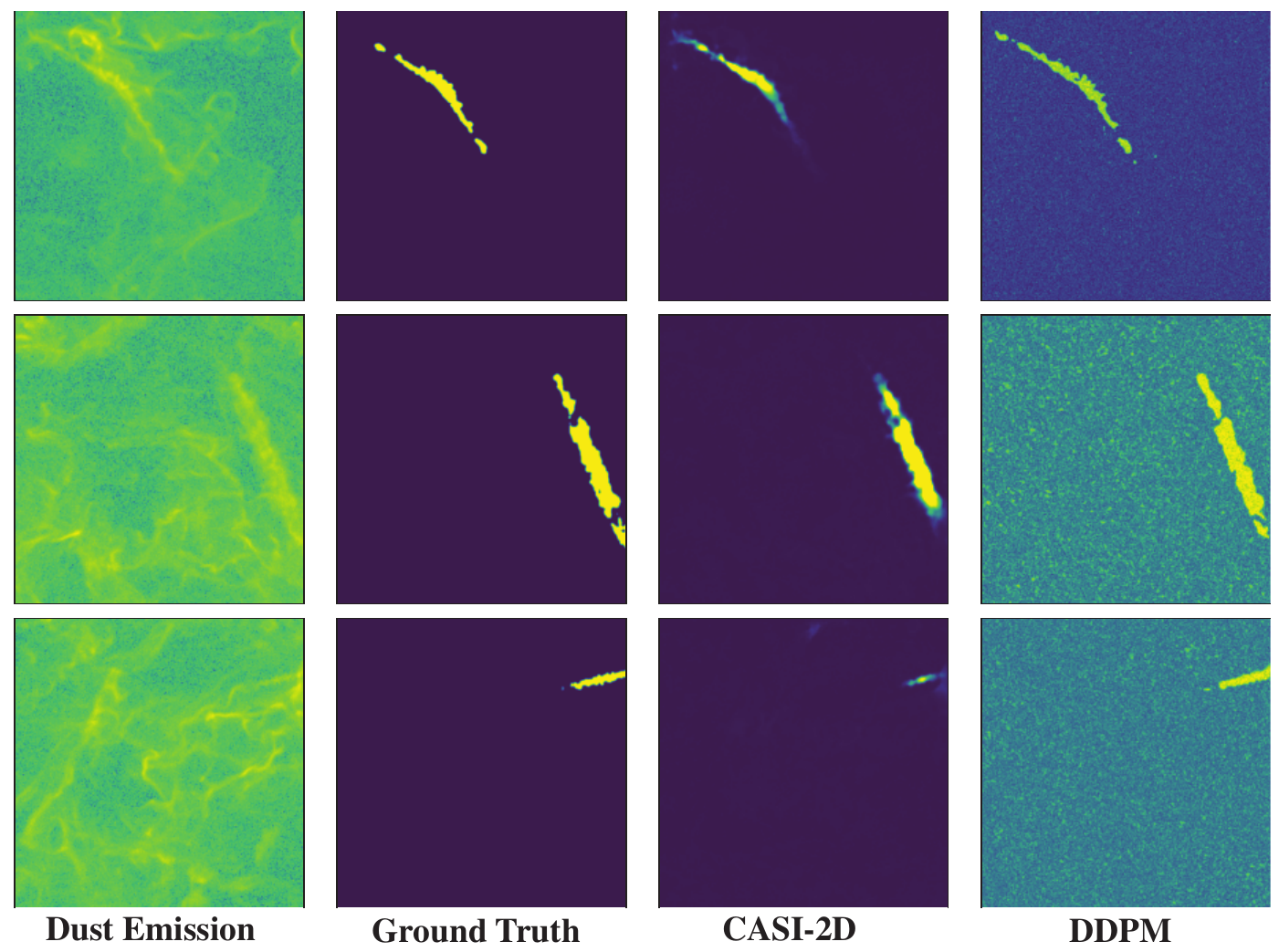}
    \caption{From \citet{2023ApJ...955..113X}, evaluation of \CASItwoD\ (a UNet architecture) and DDPM in segmenting CMR filaments in dust emission. }
    \label{fig_cmr_DDPM}
\end{figure}

However, DDPMs require significant computational resources to train and deploy. Despite this challenge, DDPMs represent a promising new approach to image segmentation, with the potential to overcome the limitations of traditional methods, especially in challenging conditions. For example, a recent study by \citet{2023ApJ...955..113X} showed that DDPMs can be used to achieve highly accurate segmentation of filamentary structures in astronomical images. Figure~\ref{fig_cmr_DDPM} presents an illustration of DDPM's application in segmenting a distinct type of filament formed through the collision-induced magnetic reconnection (CMR) mechanism in dust emission. As machine learning and image processing continue to evolve, DDPMs are poised to play an increasingly important role in image segmentation and other image-related tasks.

\subsection{Transfer Learning}

Transfer learning is a potent approach to enhance the efficacy of image segmentation tasks by capitalizing on the knowledge acquired from pre-trained neural networks. This methodology entails taking a pre-trained model, which has undergone training on an extensive dataset for a related task, and fine-tuning it for a new image segmentation objective. It is essential to recognize that transfer learning is a technique, not a fixed architecture in deep learning. Consequently, various architectures, including Unet and ViTs, can be employed in transfer learning as long as pre-trained models are accessible for use.

Transfer learning is particularly beneficial when working with limited labeled data for the specific segmentation problem at hand. In this case, the pre-trained model, often referred to as the "base model," provides a foundation of feature extraction and pattern recognition. By transferring this knowledge, the model can learn to understand and recognize common visual patterns, edges, and textures, which are essential for segmentation tasks.

The transfer learning process then fine-tunes the base model on the target dataset, which contains the images and corresponding segmentation labels specific to the new task. This fine-tuning refines the model's learned representations and adapts them to the nuances of the segmentation problem in question.

Transfer learning offers several benefits for image segmentation. First, it significantly reduces the need for a large annotated dataset, as the knowledge from the pre-trained model jumpstarts the learning process. This can be particularly advantageous in domains like medical imaging and astronomy, where obtaining labeled data can be labor-intensive and time-consuming. Second, transfer learning accelerates the training process, allowing the model to reach a desired level of accuracy with fewer iterations.

Third, transfer learning enhances the model's generalization capacity. The features extracted by the base model have been learned from extensive data, making them robust and transferable across related tasks. This adaptability proves beneficial when working with diverse images and complex backgrounds, common in fields like astronomy.

\begin{figure}[!htbp]
    \centering
    \includegraphics[width=0.99\linewidth]{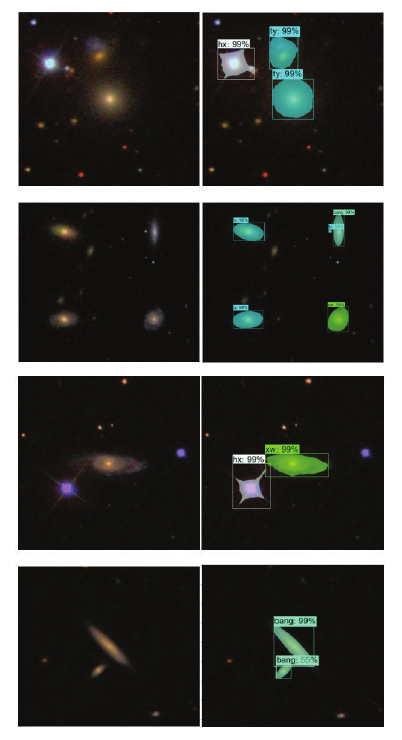}
    \caption{From \citet{GU101353372023}, an illustration of Mask R-CNN implementing transfer learning by retraining with a pre-trained model for the segmentation of various types of galaxies.}
    \label{fig_transferlearning_maskrcnn}
\end{figure}

Transfer learning offers significant advantages for image segmentation, but it is crucial to carefully select an appropriate pre-trained model, ensure compatibility between the source and target tasks, and meticulously fine-tune the model. In the domain of image segmentation, transfer learning has the potential to enhance the precision, efficiency, and practical utility of segmentation tasks across diverse domains. Notable examples include the work of \citet{2021MNRAS.508.3111M}, where a pre-trained model and transfer learning were employed to segment solar corona structures on the sun, and the study by \citep{2023Icar..39415434L}, which utilized a transfer learning approach to detect asteroid craters on the moon. In another application, \citet{GU101353372023} applied transfer learning with a Mask R-CNN architecture, starting with a model pre-trained on the Microsoft Common Objects in Context (MS COCO) dataset, and retrained it to segment galaxies. \xd{\citet{2019MNRAS.484...93D} also demonstrates the effectiveness of transfer learning by applying a pre-existing training model that classifies galaxies in one survey data to new survey data with minimal retraining using a small sample.} Figure~\ref{fig_transferlearning_maskrcnn} provides an example of the transfer learning application of Mask R-CNN in segmenting different types of galaxies. These instances underscore the versatility and effectiveness of transfer learning in addressing segmentation challenges across various scientific domains.

\subsection{Large Foundation Models}

Large foundation models have become important in image segmentation, transforming how we identify and delineate objects in images. Among large foundation models, the Segment Anything Model (SAM) represents a groundbreaking advancement, promising unparalleled flexibility and the ability to generalize to new objects and images without additional training.

\begin{figure}[!htbp]
    \centering
    \includegraphics[width=0.99\linewidth]{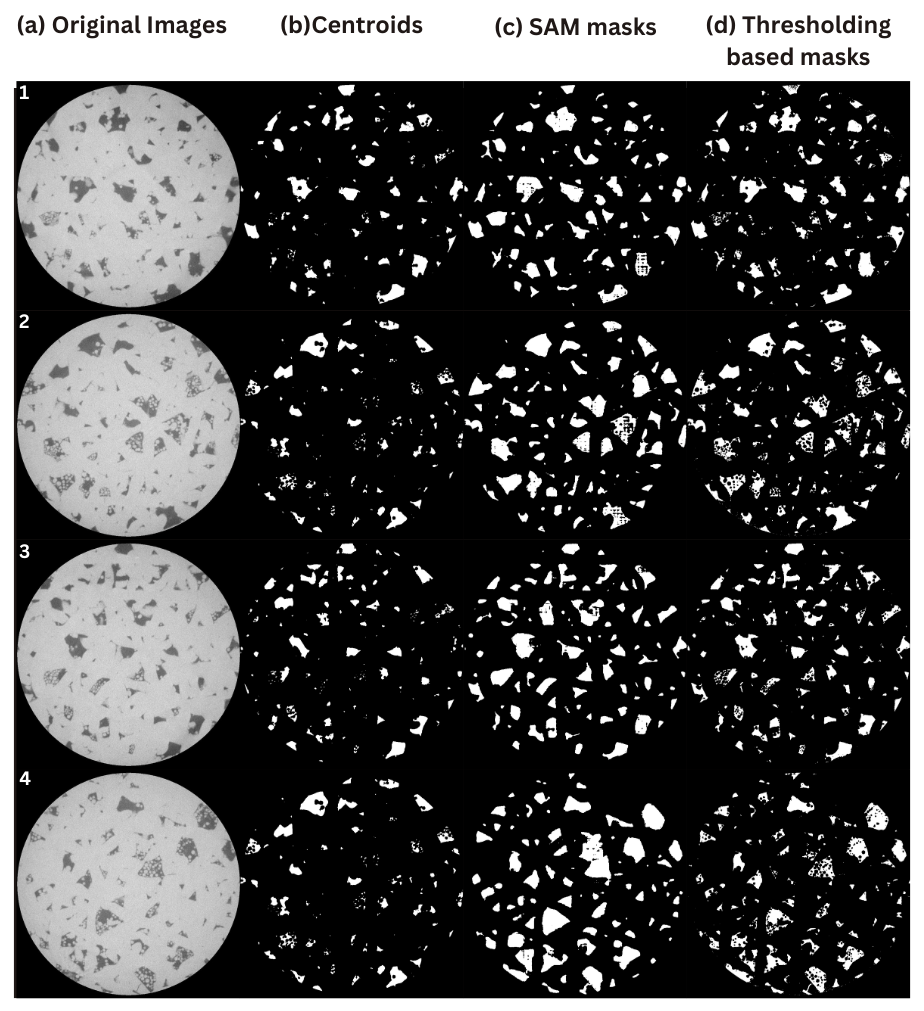}
    \caption{From \citet{2023arXiv231204063Z}, a demonstration of SAM's application in segmenting defects on a metal surface.}
    \label{fig_Final_results_SAM}
\end{figure}

Traditionally, segmentation has been divided into interactive and automatic approaches. Interactive segmentation requires user input to refine masks, while automatic segmentation requires predefined object categories and substantial training on labeled data. Both approaches have limitations, making SAM's unique promptable interface a significant advantage. SAM users can provide a variety of prompts, such as foreground/background points, bounding boxes, and masks, to perform segmentation without additional training. This flexibility makes SAM suitable for a wide range of tasks, including multimodal understanding and integration with AI systems.

SAM's remarkable adaptability also sets it apart. It can seamlessly transition between interactive and automatic segmentation, offering the best of both worlds. Moreover, SAM's training on a massive dataset of over one billion masks enables it to generalize to new objects and images with unprecedented precision.

SAM's architecture consists of three key components: a VIT-H (Vision Transformer-Huge) image encoder, a prompt encoder, and a lightweight transformer-based mask decoder. The image encoder generates an image embedding from each input image, while the prompt encoder embeds various input prompts, including interaction cues like clicks or bounding boxes. The mask decoder then predicts object masks using the image embedding and prompt embedding.

\begin{figure}[!htbp]
    \centering
    \includegraphics[width=0.99\linewidth]{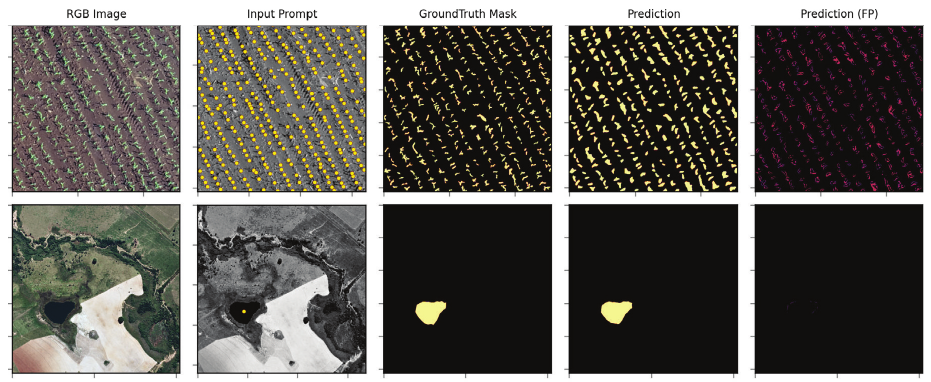}
    \caption{From \citet{OSCO2023103540}, depictions of images segmented by SAM utilizing point prompts. The initial column displays the RGB image, followed by the second column illustrating the treatment of the point prompt. The third column presents the ground-truth mask, and the fourth column exhibits the prediction result from SAM. The last column emphasizes the false-positive (FP) pixels identified in the prediction.}
    \label{fig_sam_remotesensing}
\end{figure}

SAM represents a groundbreaking large-scale model with the potential to revolutionize image segmentation, offering unparalleled flexibility and adaptability that can reshape our perception and interaction with the visual world. Its versatility is demonstrated in several pioneering applications in scientific images, such as segmenting crater structures on material surfaces \citep{gonzalez2024automatic} and defects on metal surfaces \citep{2023arXiv231204063Z}. Figure~\ref{fig_Final_results_SAM} provides an illustrative example of SAM's application in segmenting defects on a metal surface, suggesting its potential use in similar solar images for segmenting sunspots, coronal holes, and solar flares in solar observational images. \xd{Although there have not been any direct applications of SAM published in astronomy studies yet,} one notable application involves using SAM on remote sensing data observed from space to segment different terrains \citep{OSCO2023103540}. Figure~\ref{fig_sam_remotesensing} illustrates an example of SAM applied to segment different terrains in remote sensing data. This adaptability holds significant promise for astronomy, suggesting SAM's potential to segment celestial structures with unprecedented precision.

\subsection{Metrics and Evaluations}
\label{Metrics and Evaluations}



Various metrics in deep learning-based segmentation methods serve specific roles in either training or evaluation. Here, we present several widely adopted metrics in the computer vision community for tasks related to image segmentation.

\begin{itemize}
    \item  Intersection over Union (IoU): IoU is a widely employed metric for evaluating segmentation accuracy. It is calculated as the area of intersection between the predicted segmentation mask (denoted as $A$) and the ground truth mask (denoted as $B$), divided by the area of their union. Mathematically, IoU is defined as:
    \begin{align*}
        \rm{IoU} = \frac{|A \cap B|}{|A \cup B|}.
    \end{align*}
    A variant of the IoU metric is the mean IoU, which represents the average IoU computed across all target classes. \xd{Another form of generalized intersection over union, proposed by \citep{rezatofighi2019generalized}, can serve as the objective function to optimize in scenarios involving non-overlapping bounding boxes, where traditional IoU methods may be ineffective.} 

    \item Dice Coefficient: The Dice Coefficient, a prominent metric in image segmentation, particularly in medical image analysis, measures the similarity between two sets. It is commonly used to assess the agreement between a predicted segmentation mask and the ground truth mask. The Dice Coefficient is defined as:
    \begin{align*}
        \rm{Dice} = \frac{2|A \cap B|}{|A|+|B|}.
    \end{align*}
   The Dice Coefficient ranges from 0 to 1, with 0 indicating complete dissimilarity and 1 indicating a perfect match. This metric is advantageous in tasks where the class of interest is a small portion of the overall image, effectively addressing class imbalance. Similar to IoU in definition, the Dice Coefficient is useful for both training and performance evaluation.

   \item Pixel Accuracy (PA): PA is a metric used in image segmentation tasks to assess the overall accuracy of pixel-wise classification. It is calculated as the ratio of correctly classified pixels to the total number of pixels in the image. Mean Pixel Accuracy (mPA) extends this concept by averaging PA over each segmentation class. Unlike metrics such as IoU and Dice, PA and mPA are primarily used for evaluation purposes. Pixel Accuracy offers a straightforward assessment of the model's ability to correctly classify individual pixels, regardless of class labels. However, it may not be the most suitable metric for tasks with imbalanced class distributions. This is because PA treats all classes equally and may not provide a comprehensive assessment of the model's performance, especially for minority classes. In such scenarios, metrics like Intersection over Union (IoU) or Dice Coefficient are often preferred for a more nuanced evaluation.

    \item In addition to the commonly used metrics and losses discussed earlier, there are other metrics specifically designed for object segmentation in the computer vision field. Metrics like Precision, Recall, and F1 scores are frequently used for evaluation, while the cross-entropy loss is a common choice for training. These metrics are typically defined and calculated based on object class annotations, making them more suitable for multi-object semantic segmentation tasks.
    
    \end{itemize}

It is important to note that differences may exist between traditional discriminative machine learning models and generative models in their training and evaluation processes. Deep generative models aim to estimate an unknown real-world distribution and establish an effective mapping between an easy-to-sample prior (e.g., Gaussian) and the target implicit distribution. While the training goals of generative models may align, the evaluation protocols are diverse and tailored to specific application scenarios. In image generation, where fidelity matters, Fréchet Inception Distance (FID)~\citep{heusel2017fid} and Inception Score (IS)~\citep{salimans2016is} serve as widely adopted automatic metrics to assess image quality according to human perceptions. For video generation, Fréchet Video Distance (FVD)~\citep{unterthiner2019fvd}, an extension of FID, evaluates video quality. In audio generation tasks, emphasis is placed on beats and rhythms for the generated audio signals~\citep{zhu2022quantized}. In applications like image segmentation, evaluations adhere to established norms specific to the relevant downstream field of research. Despite the diversity of automated quantitative assessment methods, human inspection remains a universal benchmark for a thorough performance evaluation. \xd{Given the diversity of perspectives, individuals are likely to have varying opinions on segmentation results, and even when providing training data, different individuals may produce different masks, contributing to subjectivity in user opinions and model training. While methods like Mean Opinion Score exist to collect user opinions and rate model performance, they require substantial participation and can be labor-intensive. Only with a large and diverse group of respondents can a more converged opinion be obtained. We acknowledge that both models and individuals may exhibit bias in segmentation tasks.} Consequently, it is crucial to emphasize that all outputs from machine learning methods in image segmentation, including those from traditional segmentation methods listed in Section~\ref{Classic Segmentation Methods}, require careful human inspection before proceeding to further analysis.


\section{Conclusions}
\label{Conclusions}

In conclusion, our review of segmentation methods in astronomy has encompassed both classical techniques and the state-of-the-art machine learning approaches. These methods play a vital role in diverse scientific tasks, providing valuable insights into the complex structures present in astronomical images and data cubes.

The emergence of advanced machine learning techniques in the broader computer science community has heralded a new era of segmentation methodologies. These innovative approaches offer enhanced capabilities for precisely delineating objects in astronomical data, marking a significant advancement in the field.

As we navigate the evolving landscape of segmentation methods, we envision a future where the astronomy community fully embraces and integrates these advanced techniques into routine research endeavors. By harnessing the power of these cutting-edge segmentation methods, astronomers can benefit from more precise, reliable, and efficient segmentation outcomes. This, in turn, has the potential to alleviate astronomers from laborious manual efforts, allowing them to focus more on interpreting and understanding the intricate physical processes captured in their datasets. The synergy between advanced segmentation methods and astronomical research holds the promise of unlocking deeper insights into the mysteries of the universe.

\xd{We express our gratitude to the anonymous referees for their valuable comments and suggestions, particularly the references, which have significantly enhanced the quality of this review.}
DX acknowledges support from the Virginia Initiative on Cosmic Origins (VICO). 
YZ acknowledges support from the VisualAI lab of Princeton University.
We recognize the utilization of ChatGPT, a language model created by OpenAI using the GPT-3.5 architecture, for grammar checking in our review paper.





\bibliography{references}
\bibliographystyle{elsarticle-harv}

\end{document}